\documentclass[%
reprint,
 amsmath,amssymb,
 aps,
]{revtex4-1}

\usepackage{graphicx}
\usepackage{dcolumn}
\usepackage{bm}
\usepackage[flushleft]{threeparttable}
\usepackage{url}
\usepackage{amsmath}
\usepackage{amssymb}
\usepackage{braket}
\usepackage{multirow}
\usepackage{array}
\usepackage{url}
\newcolumntype{P}[1]{>{\centering\arraybackslash}p{#1}}

\begin{document}

\preprint{APS/123-QED}

\title{Modified Schottky emission to explain thickness dependence and slow depolarization in BaTiO$_3$ nanowires}

\author{Y. Qi,$^1$ J. M. P. Martirez,$^1$ Wissam A. Saidi,$^2$ J. J. Urban,$^3$ W. S. Yun,$^4$ J. E. Spanier,$^5$ and A. M. Rappe$^1$}
\affiliation{%
 $^1$The Makineni Theoretical Laboratories, Department of Chemistry,\\
 University of Pennsylvania, Philadelphia, PA 19104-6323 USA\\
 $^2$Department of Mechanical Engineering and Materials Science,\\
 University of Pittsburgh, Pittsburgh, PA 15261 USA\\
 $^3$The Molecular Foundry, Materials Sciences Division, \\ 
 Lawrence Berkeley National Laboratory, Berkeley, California 94720 USA \\
 $^4$Department of Chemistry, \\ 
 Sungkyunkwan University (SKKU), Suwon 440-746, Korea \\
 $^5$Department of Materials Science and Engineering, Drexel University, Philadelphia, PA 19104 USA
}%




\date{\today}

\begin{abstract}

We investigate the origin of the depolarization rates in ultrathin adsorbate-stabilized ferroelectric wires. 
By applying density functional theory calculations and analytic modeling, we demonstrate that the depolarization results from the leakage of charges stored at the surface adsorbates, 
which play an important role in the polarization stabilization. The depolarization speed varies with thickness and temperature, following several complex trends. 
A comprehensive physical model is presented, in which quantum tunneling, Schottky emission and temperature dependent electron mobility are taken into consideration.
This model simulates experimental results, validating the physical mechanism. 
We also expect that this improved tunneling-Schottky emission model could be applied to predict the retention time of polarization and the leakage current 
for various ferroelectric materials with different thicknesses and temperatures.

\end{abstract}

\pacs{Valid PACS appear here}
\maketitle


\section{Introduction}

Spontaneous electric polarization makes perovskite-based oxides of great interest for application to nonvolatile memory devices~\cite{Miller92p5999,Mathews97p238}. 
However, the polarization of ferroelectric materials may not be infinitely stable,
and retention time is one of the key factors determining the performance of memory devices in nonvolatile technology. 
The proposed reasons for the polarization instability have included the depolarization field and the leakage current~\cite{Ma02p386}, 
whose effects become more significant as the oxide film gets thinner. 
Therefore, for successful technology application, the depolarization processes of nanoscale ferroelectric oxides must be better understood.  
Here, we report a combined experimental and theoretical investigation of the depolarization process of single--crystalline BaTiO$_3$ nanowires. We
attribute the decay of polarization to the leakage of surface screening charge and propose an analytical model to explain the experimental decay rates. 

\subsection{Experimental background}
The effects of the depolarization field on the stability of ferroelectricity in ultrathin materials was
explored in BaTiO$_3$ nanowires by measuring the ferroelectric transition temperature as a 
function of the nanowire diameter in the range of $3$--$48$ nm~\cite{Spanier06p735,Urban02p1186,Yun02p447,Urban03p423}.
Positive ferroelectric domains were written perpendicular to the nanowire axis using a negative-bias voltage ($-$10 V) 
applied by a conductive scanning probe microscope cantilever tip (under ultrahigh vacuum conditions with a base pressure of ${10}^{-10}$ torr).
The time evolution of the polarized domain was then monitored via time-resolved measurements of the local electric field, 
using non-contact electrostatic force microscopy (EFM). The writing and reading processes were done at various temperatures, starting from $\approx 393$ K 
for thin nanowires ($3$--$11$ nm) and $\approx 418$ K for the thicker ones ($12$--$37$ nm).
The two sets of nanowires were progressively cooled down and retested to $\approx 308$ K and $\approx 383$ K were reached respectively. 
The Curie temperature $T_C$ was defined as the highest temperature below which the polarization signal persists for a period longer than 200 hours. 
Experiments showed that $T_C$ is inversely proportional to the diameter of the BaTiO$_3$ nanowire, in accord with standard models of depolarization field~\cite{Batra72p291}. 
At several temperatures above $T_C$,
the surface potential signals were measured during the process of polarization decay.
The magnitude of surface potential was fitted with the expression:
\begin{equation}
S(t)=S(0)e^{-k_dt}
\end{equation}
where $S(t)$ is the potential at time $t$, which is proportional to the surface screening charge,
and $k_d$ is the decay rate, with the unit s$^{-1}$~\cite{Nonnenmann09p5205}. Here, we should note that experimentally it is found that the signals decay with time approximately (but not perfectly) exponentially.
Despite the slight deviations, the decay rate $k_d$ obtained from the data fitting is still an important physical parameter in describing the polarization decay speed.
From the time evolution of the signal~\cite{Spanier06p735}, 
we see that the bright circular signal faded without expansion, which means that the depolarization is a process of leakage or tunneling, rather than diffusion, of the surface screening charge.
The observed decay rates $k_d$, which vary with nanowire thickness and temperature, are presented in Tables $\rm\uppercase\expandafter{\romannumeral1}$ and $\rm\uppercase\expandafter{\romannumeral2}$.

\begin{table}
\caption{Decay rates of ferroelectric polarization for different BaTiO$_3$ nanowire thicknesses and temperatures (thin nanowires).}
\vspace*{3 mm}
\begin{tabular}{|P{2cm}|P{1.5cm}|P{1.5cm}|P{1.5cm}|P{1.5cm}}
\cline{1-4} 
  & \multicolumn{3}{c|}{Decay Rates $k_d$ (s$^{-1}$)}   \\
\cline{2-4}
Temperature & 5 nm & 7 nm & 9 nm  \\
\cline{1-4}
393.6 K    &    0.0287$\ $ $\ $ & 0.00994$\ \,$      & 0.00368$\ \,$  & \\
\cline{1-4}
389.4 K    &      $*$            & 0.0084$\ $ $\ $    & 0.0027$\ $ $\ $    \\
\cline{1-4}
384.3 K    &    0.0232$\ $ $\ $ & 0.00485$\ \,$   &  0.00162$\ \,$   \\
\cline{1-4}
379.5 K    &    0.0161$\ $ $\ $ & 0.00248$\ \,$   &  0.000884   \\
\cline{1-4}
373.9 K    &    0.00507$\ \,$ &  0.00132$\ \,$  & 0.0001$\ $ $\ $    \\
\cline{1-4}
369.3 K    &    0.00301$\ \,$ & 0.000413   &  $*$   \\
\cline{1-4}
364.8 K    &    0.00192$\ \,$ & $*$   &  $*$   \\
\cline{1-4}
359.7 K    &    0.000916 &  $*$  & $*$    \\
\cline{1-4}
354.5 K    &    0.000287 &  $*$  & $*$    \\
\cline{1-4}
349.2 K    &    0.000267 & $*$   &  $*$   \\
\cline{1-4}
\end{tabular}
\begin{tablenotes}
\item[] $*$ Data not available
\end{tablenotes}
\end{table}
\begin{table}
\caption{Decay rates of ferroelectric polarization for different BaTiO$_3$ nanowire thicknesses and temperatures (thick nanowires).}
\vspace*{3 mm}
\begin{tabular}{|P{2cm}|P{1.5cm}|P{1.5cm}|P{1.5cm}|P{1.5cm}|}
\cline{1-5}
  & \multicolumn{4}{c|}{Decay Rates $k_d$ (s$^{-1}$)}  \\
\cline{2-5}
Temperature & 21 nm & 25 nm & 37 nm & 48 nm  \\
\hline
419 K       & 0.00333$\ \,$  &  0.00263$\ \,$  &   $*$       & 0.00171$\ \,$  \\
\hline
410 K       &    $*$      &  0.00122$\ \,$  &  0.0021$\ $ $\ $  & 0.00197$\ \,$   \\
\hline
405 K       & 0.000989 &  0.000997 &  0.00084$\ \,$ & 0.00105$\ \,$  \\
\hline
399 K       & 0.000244 &  0.00027$\ \,$  &   $*$       &    $*$       \\
\hline
395 K       & $*$ &  0.000114 &  0.000115        &   $*$       \\
\hline

\hline
\end{tabular}
\begin{tablenotes}
\item[] $*$ Data not available
\end{tablenotes}
\end{table}

The experimental data show three general trends:
$\left(1\right)$ The depolarization process is slow (several hours);
$\left(2\right)$ For any thickness, the decay rate $k_d$ increases with temperature;
$\left(3\right)$ For thin nanowires ($5$--$9$ nm), the decay rate $k_d$ changes with thickness dramatically. 
However, for thicker wires, $k_d$ stays nearly constant at different thicknesses (21, 25, 37 and 48 nm).
Our study aims at illustrating the physical essence of these trends.

\subsection{Polarization stabilization}
In recent years, many studies have investigated the dependence of polarization stabilization and leakage current on chemical environment, 
temperature, electrode material, and thickness
~\cite{Spanier06p735,Batra72p291,Stengel06p679,Szwarcman12p134112,Szwarcman14p122202,Kim05p237602,Saidi14p6711,Kolpak08p036102,Mendez-Polanco12p214107,Sai05p020101R,Kolpak06p054112,Stengel09p392,Fong06p127601,Li08p473,Wang09p047601,Koocher14p3408,Stephenson11p064107,Levchenko08p256101,He11p062905,He11p024101,Kalinin02p3816}.
From these studies, several basic principles could be drawn:
\begin{enumerate}
\renewcommand{\labelenumi}{(\theenumi)}
\item Surface polarization charge should be compensated by screening charge, in order to passivate the depolarization field and stabilize the ferroelectric distortion,
or else the polarization would become unstable. The screening charge could be stored in surface electrodes or adsorbates~\cite{Batra72p291,Stengel06p679,Szwarcman12p134112,Szwarcman14p122202,Kim05p237602,Saidi14p6711,Kolpak08p036102,Mendez-Polanco12p214107,Sai05p020101R,Kolpak06p054112};
\item The polarization state of the material may lead to preferential adsorption of certain molecules on the surface~\cite{Stengel09p392,Fong06p127601,Li08p473,Wang09p047601,Koocher14p3408,Stephenson11p064107,Levchenko08p256101}; 
\item The response of polarization with electric field or temperature is fast, but the time scale for dissipation of the surface screening charge is slow (hours or days)~\cite{He11p062905,He11p024101,Kalinin02p3816}.
\end{enumerate}

Based on the evidence above and the observations in our experiments, we propose that the physical process of depolarization in the BaTiO$_3$ nanowire experiment is as follows:
after the polarization is written, surface adsorbates on the nanowire act as an electrode that stores screening charge and stabilizes the polarization~\cite{Spanier06p735}.
For the case without external applied voltage and above $T_C$, the polarized state is not stable. 
But due to the stabilization of screening charge in the surface adsorbates, polarization in the nanowire still persists for some time.
Screening charge leaks from the top electrode (surface adsorbate) to the other side of the BaTiO$_3$ nanowire (gold substrate).
At the same time, polarization reduces along with the screening charge. This process is slow and takes hours.

In the following parts of this paper, support, analysis, and modeling of the physical processes described above are shown. 
In Section $\rm\uppercase\expandafter{\romannumeral2}$, we use density functional theory to demonstrate the role of surface adsorption in surface charge screening. 
In Section $\rm\uppercase\expandafter{\romannumeral3}$, analytical expressions describing leakage current leading to depolarization are developed. 
Finally, in Section $\rm\uppercase\expandafter{\romannumeral4}$, we present the results and discussion.

\section{DENSITY FUNCTIONAL THEORY CALCULATION}
In order to construct a theory of the depolarization process, density functional theory calculations (DFT) are carried out to assess the role of surface molecular and atomic adsorbates.
We investigate the OH molecule on BaO-terminated BaTiO$_3$ slabs, as OH is the predominant species found on oxide surfaces,
as demonstrated by both infrared spectra and ${ab}$ ${initio}$ calculations ~\cite{Spanier06p735,Noma96p5223,Wegmann04p371}.

A supercell slab method was used to describe the system. 
The supercell consists of the BaTiO$_3$ slab with 11 atomic layers, a full coverage of OH (one adsorbate per adsorption site) on the 1$\times$1 BaTiO$_3$ surface, and a vacuum of more than 20~\AA. 
The atoms are represented by norm-conserving pseudopotentials generated using the OPIUM~\cite{Opium} code with a plane-wave cutoff of 50 Ry~\cite{Al-Saidi10p155304}.

From the relaxed structure, we see that the presence of the OH adsorbates enhances ferroelectricity at the positively polarized surface
and maintains a characteristic ferroelectric displacement pattern throughout the film. 

\begin{figure}[htbp]
\includegraphics[width=8.0cm]{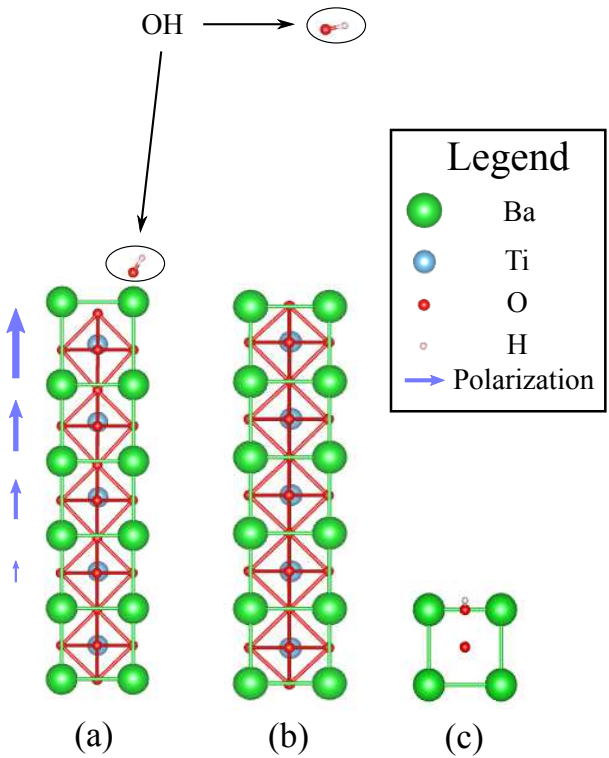}
\caption{Relaxed structures for the BaTiO$_3$/OH systems acquired from DFT calculation. Green, blue, red and gray spheres represent barium, titanium, oxygen and hydrogen atoms respectively.
(a) Polar system, OH is adsorbed at the surface; (b) Non-polar structure, OH is far away from the surface of the BaTiO$_3$ surface; (c) Top view of the OH adsorbate and the first atomic layer of BaTiO$_3$.}
\end{figure}

\begin{table}
\caption{L\"{o}wdin population in the orbitals of OH adsorbed to BaO--terminated BaTiO$_3$.}
\vspace*{3 mm}
\begin{tabular}{|P{2cm}|P{2cm}|P{2cm}|}
\hline
      & Polar    & Non-polar   \\
\hline
H $1s$ orbital  &  0.6121     &   0.5601  \\
\hline
O $2s$ orbital  &  1.7315     &   1.8138  \\
\hline
O $2p$ orbital  &  5.0719     &   4.5980  \\
\hline
Net Charge  &  -0.4155     &   0.0289  \\
\hline
\end{tabular}
\end{table}

The density of states projected onto atomic orbitals (PDOS) was calculated to characterize the charge distribution on each atom. 
Results are shown in Table $\rm\uppercase\expandafter{\romannumeral3}$, from which we see that if the BaTiO$_3$ nanowire is positively polarized,  
the hydroxyl oxygen $2p$ orbital possesses more electrons and OH is overall negatively charged.
This is a robust evidence demonstrating that surface adsorbates stabilize the polarization by holding screening charges, which has an effect similar to an electrode. 
The leakage of charge mainly from the $2p$ orbital of oxygen then results in the decay of polarization.

\section{FORMALISM OF LEAKAGE CURRENT CALCULATION}
There have been previous reports calculating the magnitude of leakage through ferroelectric films with tunneling models, Schottky emission, and the modified Schottky equation~\cite{Burstein1969tunneling,Dekkersolid,Simmons65p967}.
However, all these models come across difficulties in explaining all three depolarization rate trends mentioned in Section $\rm\uppercase\expandafter{\romannumeral1}$. B. 
A tunneling model alone cannot explain that for thick nanowires, decay rates are nearly thickness independent and for any thickness, the decay rates are strongly temperature dependent.
On the other hand, a Schottky emission model cannot account for the thin nanowire thickness dependent rates.
Here, we build a comprehensive model from the ``effective velocity'' point of view, which both accounts for the experimental results and illustrates their physical mechanisms.

The general expression for the time $\left(t\right)$ dependent leakage current $J$ can be written as
\begin{equation}
\label{ea}
-\frac{\partial{Q}\left(t\right)}{\partial{t}}=J\left(t\right)=\int{Q\left(t\right)}{n}\left(\textbf{\emph{k}}\right){v_{\rm{eff}}}\left(\textbf{\emph{k}}\right){d}^3\textbf{\emph{k}}
\end{equation}
${n}\left(\textbf{\emph{k}}\right)$ is the probability that an electron possesses a wave vector between $\textbf{\emph{k}}$ and $\textbf{\emph{k}}+{d}\textbf{\emph{k}}$.
${v_{\rm{eff}}}\left(\textbf{\emph{k}}\right)$ is the effective velocity along the $z$ direction, which is normal to the surface of the BaTiO$_3$ nanowire, for the electrons with wave vector $\textbf{\emph{k}}$. 
$Q$ is the amount of extra charge (compared with neutral OH) stored in the adsorbate. 
In the following subsections, we define the parameterization of equation~\eqref{ea}. 

\subsection{Wave vector distribution of electrons in adsorbate}
Unlike in traditional metal electrodes, electrons occupying orbitals localized on the OH adsorbates cannot be treated as a free electron gas, 
and the wave vector distribution does not follow Fermi-Dirac statistics.
Instead, the wave vector spectrum can be estimated from Bessel-Fourier transformation of the $2p$ orbital of oxygen, since screening charge is mainly associated with this atomic orbital.     
Here, the radial part of the $2p$ orbital of the oxygen is represented by a double--zeta function~\cite{Pyykko84p4892}.
\begin{equation}
\label{e3}
\begin{aligned}
&\phi_{2p}\left(\textbf{\emph{r}}\right)=R\left(r\right)Y_{10}\left(\theta_r,\phi_r\right)      \\  
&=\left(c_1\sqrt{\frac{\left(2z_1\right)^5}{4!}}re^{-z_1r}+c_2\sqrt{\frac{\left(2z_2\right)^5}{4!}}re^{-z_2r}\right)Y_{10}\left(\theta_r,\phi_r\right)        \\
&=\left(c_1^{\prime}re^{-z_1r}+c_2^{\prime}re^{-z_2r}\right)Y_{10}\left(\theta_r,\phi_r\right)        \\
\end{aligned}
\end{equation}
\begin{equation}
\begin{aligned}
\label{a7}
&\phi_{2p}\left(\textbf{\emph{k}}\right)=\frac{1}{\left(2\pi\right)^{3/2}}\int{e^{-i\textbf{\emph{k}}\cdot\textbf{\emph{r}}}}\phi_{2p}\left(\textbf{\emph{r}}\right)d^3{\textbf{\emph{r}}} \\
&=-\sqrt{\frac{8}{\pi}}4iY_{10}\left(\theta_k,\phi_k\right)\left[\frac{c_1^{\prime}z_1k}{\left(z_1^2+k^2\right)^3}+\frac{c_2^{\prime}z_2k}{\left(z_2^2+k^2\right)^3}\right] \\
\end{aligned}
\end{equation}
\begin{multline}
{n}\left(\textbf{\emph{k}}\right)=\left|\phi_{2p}\left(\textbf{\emph{k}}\right)\right|^2                         \\
=\frac{128}{\pi}\left|Y_{10}\left(\theta_k,\phi_k\right)\right|^2\left[\frac{c_1^{\prime}z_1k}{\left(z_1^2+k^2\right)^3}+\frac{c_2^{\prime}z_2k}{\left(z_2^2+k^2\right)^3}\right]^2
\end{multline}
$Y_{10}$ is the spherical harmonic for $l=1$ and $m=0$. $c_{1,2}$ and $z_{1,2}$ are the parameters in the double--zeta function, acquired from previous reference~\cite{Pyykko84p4892}. 
$c_1^{\prime}$ and $c_2^{\prime}$ are reduced coefficients taking the normalization factors $\sqrt{\frac{\left(2z_1\right)^5}{4!}}$ and $\sqrt{\frac{\left(2z_2\right)^5}{4!}}$ into consideration.
$\theta_{r,k}$ and $\phi_{r,k}$ are the angles between the directions of $\textbf{\emph{r}}$, $\textbf{\emph{k}}$ and the axes in spherical coordinates.
$\phi_{2p}\left(\textbf{\emph{r}}\right)$ and $\phi_{2p}\left(\textbf{\emph{k}}\right)$ are expressions for the oxygen $2p$ wavefunction in coordinate and 
wave vector representations. In this way, we obtain an analytical expression for ${n}\left(\textbf{\emph{k}}\right)$.
Here, we should note that in part $\rm\uppercase\expandafter{\romannumeral2}$, we used pseudo wavefunctions in the DFT calculations;
compared with all--electron wavefunctions, 
pseudo wavefunctions have lower high--$k$ components on purpose to limit the number of plane waves used~\cite{Rappe90p1227}.
This does not affect the accuracy of charge leakage rate calculation. This is because the DFT calculation is used only to illustrate the role of screening charge and the mechanism of the depolarization process.
The wave-vector distribution in this model is derived from the double-zeta function as described above. 
Additionally, in the later discussion, we will also include the fact that high--speed electrons lose their initial momentum quickly and drift under the effect of the electric field.
Therefore, an underestimation of the high--$k$ components has little effect on the charge dissipation speed.     

\subsection{Effective velocity}
Here, we present an effective velocity model of the charge dissipation.
The band diagram of the OH/BaTiO$_3$/Au substrate system is shown in FIG.2. 
\begin{figure}[htbp]
\includegraphics[width=8.0cm]{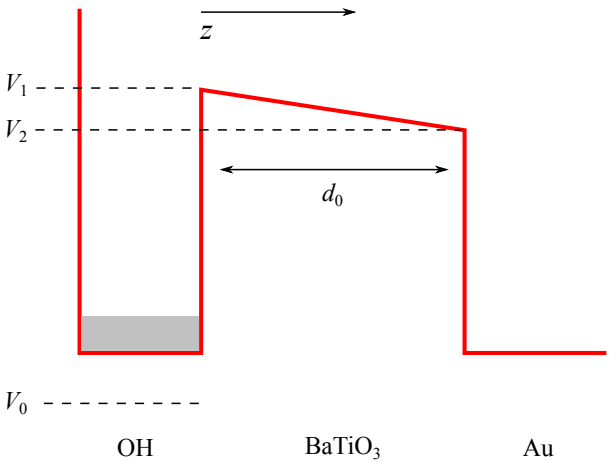}
\caption{Band diagram of the Au electrode/oxide insulator/adsorbate electrode system.}
\end{figure}
We only consider the effective velocities of the electrons with the wave vector pointing toward the nanowire $\textbf{\emph{k}}\cdot{\bm{\hat{z}}}>0$. 
Otherwise, the electron does not contribute to the leakage current $J$ and its effective velocity $v_{\rm{eff}}=0$. 
For the electron moving toward the nanowire, the expression for effective velocity varies depending on whether the energy of the electron is higher than that of BaTiO$_3$ conduction band edge.
The potential energy of an electron affiliated with the hydroxyl molecule oxygen $2p$ orbital is set as $V_0$. 
The total energy for this electron can be written as the sum of its kinetic energy $T\left(\textbf{\emph{k}}\right)$ and $V_0$:
\begin{equation}
E\left(\textbf{\emph{k}}\right)=T\left(\textbf{\emph{k}}\right)+V_0=\frac{{\hbar}^2\left|\textbf{\emph{k}}\right|^2}{2m_0}+V_0.
\end{equation}
As shown in FIG. 2, the conduction band for the BaTiO$_3$ nanowire is not flat, 
and the slope equals the electronic charge $e$ times the electric field $E_{\rm{fe}}$ inside the ferroelectric nanowire.
Therefore,
\begin{equation} 
V_2=V_1-eE_{\rm{fe}}d_0,
\end{equation}
For the case $E\left(\textbf{\emph{k}}\right)<V_2$, the mechanism that governs the electron movement is quantum tunneling.
The electron tunnels from the surface adsorbate hydroxyl to the gold electrode at the other side, 
through the BaTiO$_3$ nanowire as an energy barrier.
Around $T_C$, the dielectric constant of BaTiO$_3$ is large and $E_{\rm{fe}}$ is small. 
Therefore, the conduction band is nearly flat and we use the approximation that the energy barrier is a cuboid with the height $V_2$.
The transmission coefficient $P\left(\textbf{\emph{k}}\right)$, which is also the probability of one electron with wave vector $\textbf{\emph{k}}$ penetrating the barrier,
could be expressed~\cite{Mcquarrie1997physical}:
\begin{multline}
\label{a8}
P\left(\textbf{\emph{k}}\right)=  \\
\cfrac{4}{4+
\cfrac{\left[m_0E+m^*\left(V_2-E\right)\right]^2}{m_0m^*E\left(V_2-E\right)}\sinh^2\left[\cfrac{2m^*d^2\left(V_2-E\right)}{\hbar^2}\right]^{1/2}}
\end{multline}
where $d$ is the length of the barrier. 
$m^*$ is the effective mass of electrons in BaTiO$_3$.
Assuming that the angle between the incident direction of the electron and the normal direction of the BaTiO$_3$ nanowire is $\theta$, as shown in FIG. 3,
$d$ could be expressed as 
\begin{equation} 
d=\frac{d_0}{\cos\theta}
\end{equation}

\begin{figure}[htbp]
\includegraphics[width=7.0cm]{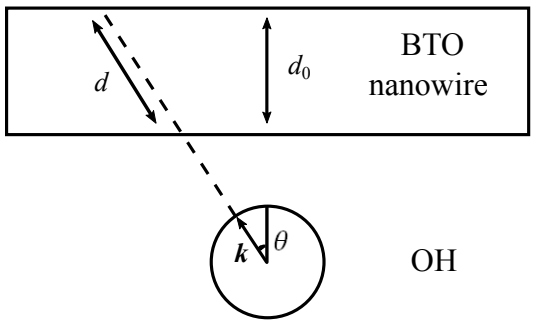}
\caption{Relationship of barrier thickness and direction of wave vector.}
\end{figure}

In this tunneling process, an electron with an initial velocity $v=\frac{\hbar\textbf{\emph{k}}}{m_0}$ has a probability $P\left(\textbf{\emph{k}}\right)$ of
passing the energy barrier with the thickness $d$. Thus, the effective velocity could be expressed as
\begin{equation} 
v_{\rm{eff}}\left(\textbf{\emph{k}}\right)=v{P\left(\textbf{\emph{k}}\right)}=\frac{\hbar\textbf{\emph{k}}\cdot{\bm{\hat{z}}}}{m_0}P\left(\textbf{\emph{k}}\right)
=\frac{\hbar{k}{\cos{\theta}}}{m_0}P\left(\textbf{\emph{k}}\right)
\end{equation}

When the hydroxyl electron energy is higher than that of the BaTiO$_3$ conduction band edge $E\left(\textbf{\emph{k}}\right)>V_1$, 
the electron is not classically forbidden to enter the BaTiO$_3$ nanowire.
%
%
%
%
However, very high velocity electrons could have very short mean free paths. 
Moreover, in the
BaTiO$_3$ crystal, which is an insulator and possesses a small mean free path~\cite{Zafar98p3533}, the electron loses its initial momentum quickly and drifts under the effect of electric field $E_{\rm{fe}}$ inside the ferroelectric nanowire:
\begin{equation} 
v_{\rm{eff}}={\mu}E_{\rm{fe}},
\end{equation}
where $\mu$ is the electronic mobility, which depends on temperature, intermediate energy levels and lattice vibrations. 
Lattice vibrations have a significant influence on the drift mobility due to electron-phonon interactions~\cite{Ziman1960electronics,Low55p414,Schultz59p526,Wemple69p547}, in which an electron may lose or gain energy in a collision. 
This effect will be discussed in the following subsection.

For the final case, $V_2<E\left(\textbf{\emph{k}}\right)<V_1$, the electron first tunnels through an energy barrier and then drifts in the conduction band. 
The effective velocity is given by
\begin{equation} 
v_{\rm{eff}}={\mu}E_{\rm{fe}}P\left(\textbf{\emph{k}}\right).
\end{equation}
For this case, the energy barrier goes to zero within the nanowire. 
At this the condition, the Wentzel--Kramers--Brillouin (WKB)~\cite{Spigler95p567} approximation fails.
Since $E\left(\textbf{\emph{k}}\right)$ is close to $V_1$, we approximate $P\left(\textbf{\emph{k}}\right)$ to 1.

Even though the approximations in the first and last cases overestimate the tunneling current, 
later we will point out that despite this, the tunneling current is still a negligible contribution to the overall current.

\subsection{Electric Properties of BaTiO$_3$ Nanowire}
The relationship of surface charge density $Q$, electric displacement $D_{\rm{fe}}$ and electric field $E_{\rm{fe}}$ though the insulator is given by dielectric constant $\epsilon\left(T\right)$~\cite{Szwarcman12p134112}:
\begin{equation} 
Q\left(t\right)=D_{\rm{fe}}=\epsilon\left(T\right){\epsilon_0}{E_{\rm{fe}}}
\end{equation}

According to the Lyddane-Sachs-Teller~\cite{Lyddane41p673} relation, the temperature dependent static dielectric constant is expressed as:
\begin{equation} 
\frac{\epsilon\left(T\right)}{\epsilon_\infty}=\frac{\prod_i{\omega_{L_i}^2}\left(T\right)}{\prod_i{\omega_{T_i}^2\left(T\right)}}
\end{equation}
$\omega$ is the frequency of lattice vibration and $T_{i}$ and $L_{i}$ represent transverse and longitudinal modes. 
Approaching $T_C$, a transverse optical mode $\omega_{\rm{TO}}$ becomes ``soft'', indicating the occurrence of phase transition~\cite{Shirane67p234,Rabe07,Singh96p176}:
\begin{equation} 
\omega_{\rm{TO}}^2\approx{A}\left(T-T_C\right)
\end{equation}
\begin{equation} 
{\epsilon\left(T\right)}^{-1}\propto\omega_{\rm{TO}}^2={A^{\prime}}\left(T-T_C\right)
\end{equation}
$A$ and $A^{\prime}$ are constants, and for BaTiO$_3$ bulk~\cite{Benedict59p1091,Barker66p391}, $T_C\approx{393}$ K. However, for a thin film or nanowire, if polarization exists, a depolarization field
is induced due to the incomplete charge compensation of polarization charge~\cite{Wurfel73p5126,Mehta73p3379,Kim05p237602}. 
The depolarization field, which is anti--parallel to polarization, becomes significant as the thickness decreases. 
The depolarization field applies an electric force on the ions, and as a consequence, 
the soft vibrational mode is hardened. 
With the depolarization field effect included, the temperature dependent soft mode frequency for thin films should be rewritten as
\begin{equation} 
\label{e6}
\omega_{\rm{TO}}^2\approx{A}\left(T-T_C^{\prime}\right)
\end{equation}
\begin{equation} 
\label{e4}
{\epsilon\left(T\right)}^{-1}={A^{\prime}}\left(T-T_C^{\prime}\right)
\end{equation}
and 
\begin{equation} 
T_C^{\prime}<T_C
\end{equation}
The depolarization field effect hardens the soft mode, increases the vibrational frequency, suppresses the ferroelectricity and lowers the $T_C$.
The values of $T_C^{\prime}$ for different thicknesses could be inferred from $T_C$, which were previously experimentally measured~\cite{Spanier06p735}. 

Previous studies illustrated that the electronic mobility in BaTiO$_3$ varies over a large range (10$^{-3}$--10$^{-1}$ cm$^2$/V--s)~\cite{Berglund67p358,Boyeaux79p545,Wemple69p547}.
The variation of mobility could be attributed to the different trap levels, types and concentrations from the fabrication process. 
Despite the uncertainty, there is a general rule that electron mobility is determined by the optical phonons~\cite{Ziman1960electronics,Low55p414,Schultz59p526,Wemple69p547}.   


Many studies in recent years pointed out that scattering by the soft transverse optical mode is the primary factor affecting electron drift~\cite{Wemple69p547,Fridkin73p71,Vinetskii70p23}. In this case~\cite{Vinetskii70p23}:
\begin{equation} 
\mu=\frac{f\left(T\right)}{\epsilon\left(T\right)}\propto{f}\left(T\right){\omega_{\rm{TO}}^2}
\end{equation}
and $f\left(T\right)$ depends weakly on $T$ for BaTiO$_3$~\cite{Vinetskii70p23,Fridkin73p71}. For $T>T_C^{\prime}$
\begin{equation} 
\label{e2}
\mu\approx{B}\omega_{\rm{TO}}^2=B^{\prime}\left(T-T_C^{\prime}\right)
\end{equation}
$B$ and $B^{\prime}$  are constants.
An explicit explanation about the soft mode dependent mobility was proposed in Ref.~\cite{Wemple69p547}. 
In brief, according to thermodynamics, the mean squared polarization fluctuation $\delta{P}$ is related to the dielectric constant as~\cite{burgess1965fluctuation}. 
\begin{equation}
\langle{\delta{P}^2}\rangle=k_BT\epsilon/V\propto1/{\omega_{\rm{TO}}^2}
\end{equation}
$k_B$ is Boltzmann constant and $V$ is the volume. For the case $T>T_C^{\prime}$, a high soft mode frequency means a small dielectric constant, large mean--square TO phonon amplitude and polarization fluctuation. 
Besides, in a small region with an approximately uniform polarization fluctuations, shifts in the conduction band edge $\Delta\mathcal{E}_c$ is~\cite{shockley1953electrons}
\begin{equation} 
\Delta\mathcal{E}_c={\rm{const}}+\beta\delta{P}^2
\end{equation}
$\beta$ is the polarization--potential parameter, which has a typical value $\beta\approx2$ eV m$^4$/C$^2$.
In the nanowire, the low soft mode frequency leads to inhomogeneity in $\Delta\mathcal{E}_c$, which then results to locally different effective masses and electronic energies.
Therefore, an electron traveling through the wire scatters more. 
This is similar to an electron traveling on a curved path and harder to accelerate.
A lower soft mode frequency means more scattering, a shorter relaxation time, $\tau_e$, and a smaller electron mobility, since electron mobility is given by~\cite{kittel1986introduction}
\begin{equation} 
\mu=\frac{e\tau_e}{m^*}
\end{equation}
Thus, in this simulation, we use the empirical expression of electron mobility as shown in equation~\eqref{e2}.

\section{results and discussions}

With the expressions of effective velocity deduced in different wave vector range, 
\begin{equation}
\label{e5}
v_{\rm{eff}}\left(\textbf{\emph{k}}\right) = \left\{ \begin{array}{rl}
& \frac{\hbar\textbf{\emph{k}}}{m}{\cdot}P\left(\textbf{\emph{k}}\right) \ \, \mbox{\quad if $E\left(\textbf{\emph{k}}\right)<V_2$, $\textbf{\emph{k}}\cdot{\bm{\hat{z}}}>0$} \\
& \\
& {\mu}E_{\rm{fe}}P\left(\textbf{\emph{k}}\right) \quad \mbox{if $V_2<E\left(\textbf{\emph{k}}\right)<V_1$, $\textbf{\emph{k}}\cdot{\bm{\hat{z}}}>0$} \\
\\
& {\mu}E_{\rm{fe}}     \quad  \ \,   \mbox{\qquad if $E\left(\textbf{\emph{k}}\right)>V_1$, $\textbf{\emph{k}}\cdot{\bm{\hat{z}}}>0$} \\
& \\
& 0           \quad    \quad     \quad   \mbox{\quad \quad if $\textbf{\emph{k}}\cdot{\bm{\hat{z}}}<0$,}
       \end{array} \right.
\end{equation}

\noindent we could calculate the time evolution of surface charge with equation~\eqref{ea}.

The expression of current density was shown in equation $\left(1\right)$. 
In the calculation of charge density evolution with time,
the initial charge density used in the simulation is $Q\left(0\right)=0.26$ C/m$^2$~\cite{Al-Saidi10p155304}, which is the spontaneous polarization for the tetragonal phase of BaTiO$_3$.
The time window is selected as $10^4$ s, which is long enough to demonstrate the general trend of time evolution of surface charge in adsorbate OH.
Other parameters involved in the presented tunneling and modified Schottky model are listed in TABLE $\rm\uppercase\expandafter{\romannumeral4}$. 
Most of the parameters are from previous references.  
The difference of the potential energy of electrons in the adsorbate oxygen $2p$ orbital $\left(V_0\right)$ and in the conduction band formed by titanium $3d$ orbitals $\left(V_1\right)$ 
is estimated from the band gap of BaTiO$_3$ (3.20 eV)~\cite{Cardona65p651}.
This is a good approximation because the adsorbate oxygen $2p$ orbital is approximately at the same level with the valence band formed by oxygen $2p$ orbitals of the nanowire. 
\begin{equation} 
V_1-\overline{{E}\left(\textbf{\emph{k}}\right)}\approx3.20\ \rm{eV}
\end{equation}
where $\overline{{E}\left(\textbf{\emph{k}}\right)}$ is the average energy of electrons in the adsorbate oxygen $2p$ orbital, and 
\begin{equation} 
\overline{{E}\left(\textbf{\emph{k}}\right)}=\overline{{T}\left(\textbf{\emph{k}}\right)}+V_0
\end{equation} 
\begin{equation} 
\overline{{T}\left(\textbf{\emph{k}}\right)}=\int_0^\infty{n}\left(\textbf{\emph{k}}\right)\frac{\hbar^2{\textbf{\emph{k}}}^2}{2m_0}{d}^3\textbf{\emph{k}}=68.28\ \rm{eV}
\end{equation}
\begin{equation} 
V_1-V_0=V_1-\overline{{E}\left(\textbf{\emph{k}}\right)}+\overline{{T}\left(\textbf{\emph{k}}\right)}=71.48\ \rm{eV}
\end{equation}

In the modeling, the surface charge density $Q$ decays with time $t$, but not exactly exponentially. We fit each $Q$ \emph{vs.} $t$ curve with an exponential function by the least squares fitting method.
In this way, we obtain decay constants from this model that can be compared with experimental ones.
$B^{\prime}$ of equation~\eqref{e2} is the only parameter calculated by fitting the data in experiments to the decay rates calculated by this analytical model, rather than from any references or ${ab}$ ${initio}$ calculation.

\begin{table}[htbp]
\caption{Parameters involved in the presented tunneling and modified Schottky model of nanowire depolarization.}
\begin{center}
\begin{tabular}{lll} 
\hline
 Parameter & Description & Value \\ 
\hline
 $T_C^{\prime}$ (5 nm)      & $T_C$ for 5 nm nanowire$^a$                          &   340.3 K  \\
 $T_C^{\prime}$ (7 nm)      & $T_C$ for 7 nm nanowire$^a$                          &   355.2 K  \\
 $T_C^{\prime}$ (9 nm)      & $T_C$ for 9 nm nanowire$^a$                          &   367.9 K  \\
 $T_{C}$           & $T_C$ for thick nanowires$^a$                          &   391 K  \\
 $c_1$               & Parameter in equation~\eqref{e3}$^b$                           &   0.72540  \\
 $c_2$               & Parameter in equation~\eqref{e3}$^b$                           &   0.35173  \\
 $z_1$               & Parameter in equation~\eqref{e3}$^b$                           &   1.62807 bohr$^{-1}$ \\
 $z_2$               & Parameter in equation~\eqref{e3}$^b$                           &   3.57388 bohr$^{-1}$ \\
 $m^{*}$             & Effective electronic mass                       &   6.5 $m_0$  \\
 $V_1-V_0$           & Energy barrier$^d$                                               &   71.48 eV  \\
 $A^{\prime}$           &          Defined in equation~\eqref{e4} $^e$                  &  7.84$\times{10}^{-6}$ K$^{-1}$ \\
  $B^{\prime}$           &          Defined in equation~\eqref{e2} $^f$                  &  5.025$\times{10}^{-6}$\\
\hline
\end{tabular}
\end{center}
\begin{tablenotes}
\item[] $^a$ Reference~\cite{Spanier06p735}
\item[] $^b$ Reference~\cite{Pyykko84p4892}
\item[] $^c$ Reference~\cite{Berglund67p358}
\item[] $^d$ From the band gap estimation
\item[] $^e$ Reference~\cite{Barker66p391}
\item[] $^f$ The unit is cm$^2$/VsK
\end{tablenotes}

\end{table}

\begin{figure}[htbp]
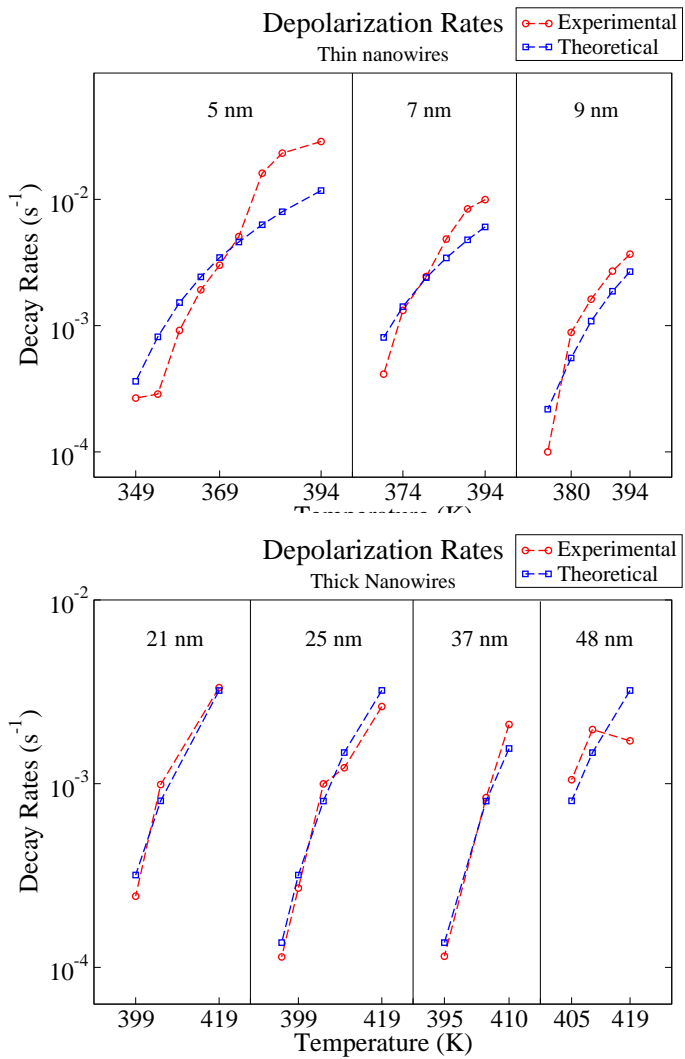

\includegraphics[width=9.0cm]{thin.eps}
\includegraphics[width=9.0cm]{thick.eps}
\caption{Comparision of experimental and theoretical depolarization rates calculated from the presented model.}
\end{figure}
The decay rates calculated from this model and experimental data are shown in FIG. 4. Experimental data are marked with red circles and theoretically calculated ones are blue squares.
Symbols for decay rates of a given thickness are connected with dashed lines, to demonstrate their temperature dependence.
From the comparison, it could been seen that this model not only simulates the experimental results to a good extent, but also sheds light on the general rules of ferroelectric leakage current.

For nanowires with any thicknesses, calculations based on equations~\eqref{ea} and~\eqref{e5} demonstrate that tunneling current provides a negligible part of the total leakage current.
Even for the thinnest nanowires (5 nm) with a surface charge density $Q\left(0\right)=0.26$ C/m$^2$, the tunneling current is around 10$^{-6}$ A/m$^{2}$ or less, generally no more than 1 percent of the total leakage current.
Emission of electrons with energies higher than the barrier contributes nearly all the current.
The fact that tunneling, whose rate is thickness dependent, plays a minor role is consistent with the experimental finding that decay rates are thickness independent for thick nanowires~\cite{Dietz97p2359}.  

For thin nanowires, at a certain temperature which is above $T_C$, the soft mode frequency decreases with thickness due to the depolarization field effect. 
A thinner nanowire corresponds to a higher soft mode frequency and a smaller dielectric constant, as shown in equation~\eqref{e6}. 
Thus, the electric field through a thinner nanowire is larger and accelerates the electron emission. 
Meanwhile, a higher transverse optical mode frequency results in a faster electron drift mobility. 
This makes the leakage current and polarization decay constant significantly thickness--dependent for thin wires. 
For thick nanowires, the soft mode frequency at a certain temperature approaches that of the bulk BaTiO$_3$ crystal. In this range, thickness affects the electric field, electron mobility, and leakage current little.
It is worth mentioning that the conclusion, that leakage current depends on thicknesses only for thin nanowires or films, is consistent with previous studies~\cite{Gruverman09p3539,Garcia09p81,Dietz97p2359}.

In this study, the surface adsorbate OH plays the role of top electrode, which is the source of nonequilibrium charge carriers. 
This modified Schottky model, simulating leakage current in ferroelectric oxides, differs from the traditional Schottky emission model in many aspects.
The electrons are localized in a 2$p$ orbital, and their wave vector distribution is now not based on Fermi-Dirac statistics.
Different from the traditional Schottky emission model, in which the thermal population of electrons leads to the temperature dependence of leakage current, 
temperature dependent leakage current is attributed to change of electric field through the nanowire and electronic mobility accompanied with the hardening (or softening)
of the transverse optical mode in this model.

\section{Conclusion}

In summary, the depolarization process of BaTiO$_3$ nanowires has been studied by both experiment and first principles calculation. 
We investigated the mechanisms which govern the polarization decay and drew several principles. 
A new proposed theoretical model, which combines molecular orbital theory, quantum tunneling and the modified Schottky equation, could explain successfully the general trends 
in the temperature and nanowire thickness dependent decay rates. Our study demonstrates that the surface adsorbate plays a significant role in stabilizing ferroelectricity and that
depolarization is a process of charge leaking from the hydroxyl surface adsorbate to the gold substrate.

\section*{ACKNOWLEDGMENTS}

Y.Q. was supported by the National Science Foundation, under Grant Number CMMI-1334241. 
J.M.P.M. was supported by the Department of Energy Office of Basic Energy Sciences, under grant number DE-FG02-07ER15920.
W.A.S was supported by the Office of Naval Research, under Grant No. N00014-12-1-1033. 
J.J.U. acknowledges support from the Molecular Foundry, which is supported by the Office of Science, 
Office of Basic Energy Sciences, at the U.S. Department of Energy (DOE), Contract No. DE-AC02-05CH11231.
W.S.Y. acknowledges support from the National Research Foundation of Korea, under Grant Number NRF-2012-0009565.
J.E.S was supported by the National Science Foundation, under Grant Number DMR-1124696.
A.M.R. was supported by the National Science Foundation, under Grant Number DMR-1124696.
Computational support was provided by the High-Performance Computing Modernization Office of the Department of Defense and the National Energy Research Scientific Computing Center.
We would like to acknowledge professor Hongkun Park for his guidance in the synthesis of the BaTiO$_3$ nanowires.
\bibliography{rappecites}

\begin{thebibliography}{63}%
\makeatletter
\providecommand \@ifxundefined [1]{%
 \@ifx{#1\undefined}
}%
\providecommand \@ifnum [1]{%
 \ifnum #1\expandafter \@firstoftwo
 \else \expandafter \@secondoftwo
 \fi
}%
\providecommand \@ifx [1]{%
 \ifx #1\expandafter \@firstoftwo
 \else \expandafter \@secondoftwo
 \fi
}%
\providecommand \natexlab [1]{#1}%
\providecommand \enquote  [1]{``#1''}%
\providecommand \bibnamefont  [1]{#1}%
\providecommand \bibfnamefont [1]{#1}%
\providecommand \citenamefont [1]{#1}%
\providecommand \href@noop [0]{\@secondoftwo}%
\providecommand \href [0]{\begingroup \@sanitize@url \@href}%
\providecommand \@href[1]{\@@startlink{#1}\@@href}%
\providecommand \@@href[1]{\endgroup#1\@@endlink}%
\providecommand \@sanitize@url [0]{\catcode `\\12\catcode `\$12\catcode
  `\&12\catcode `\#12\catcode `\^12\catcode `\_12\catcode `\%12\relax}%
\providecommand \@@startlink[1]{}%
\providecommand \@@endlink[0]{}%
\providecommand \url  [0]{\begingroup\@sanitize@url \@url }%
\providecommand \@url [1]{\endgroup\@href {#1}{\urlprefix }}%
\providecommand \urlprefix  [0]{URL }%
\providecommand \Eprint [0]{\href }%
\providecommand \doibase [0]{http://dx.doi.org/}%
\providecommand \selectlanguage [0]{\@gobble}%
\providecommand \bibinfo  [0]{\@secondoftwo}%
\providecommand \bibfield  [0]{\@secondoftwo}%
\providecommand \translation [1]{[#1]}%
\providecommand \BibitemOpen [0]{}%
\providecommand \bibitemStop [0]{}%
\providecommand \bibitemNoStop [0]{.\EOS\space}%
\providecommand \EOS [0]{\spacefactor3000\relax}%
\providecommand \BibitemShut  [1]{\csname bibitem#1\endcsname}%
\let\auto@bib@innerbib\@empty
\bibitem [{\citenamefont {Miller}\ and\ \citenamefont
  {McWhorter}(1992)}]{Miller92p5999}%
  \BibitemOpen
  \bibfield  {author} {\bibinfo {author} {\bibfnamefont {S.}~\bibnamefont
  {Miller}}\ and\ \bibinfo {author} {\bibfnamefont {P.}~\bibnamefont
  {McWhorter}},\ }\href@noop {} {\bibfield  {journal} {\bibinfo  {journal}
  {Journal of applied physics}\ }\textbf {\bibinfo {volume} {72}},\ \bibinfo
  {pages} {5999} (\bibinfo {year} {1992})}\BibitemShut {NoStop}%
\bibitem [{\citenamefont {Mathews}\ \emph {et~al.}(1997)\citenamefont
  {Mathews}, \citenamefont {Ramesh}, \citenamefont {Venkatesan},\ and\
  \citenamefont {Benedetto}}]{Mathews97p238}%
  \BibitemOpen
  \bibfield  {author} {\bibinfo {author} {\bibfnamefont {S.}~\bibnamefont
  {Mathews}}, \bibinfo {author} {\bibfnamefont {R.}~\bibnamefont {Ramesh}},
  \bibinfo {author} {\bibfnamefont {T.}~\bibnamefont {Venkatesan}}, \ and\
  \bibinfo {author} {\bibfnamefont {J.}~\bibnamefont {Benedetto}},\ }\href@noop
  {} {\bibfield  {journal} {\bibinfo  {journal} {Science}\ }\textbf {\bibinfo
  {volume} {276}},\ \bibinfo {pages} {238} (\bibinfo {year}
  {1997})}\BibitemShut {NoStop}%
\bibitem [{\citenamefont {Ma}\ and\ \citenamefont {Han}(2002)}]{Ma02p386}%
  \BibitemOpen
  \bibfield  {author} {\bibinfo {author} {\bibfnamefont {T.}~\bibnamefont
  {Ma}}\ and\ \bibinfo {author} {\bibfnamefont {J.-P.}\ \bibnamefont {Han}},\
  }\href@noop {} {\bibfield  {journal} {\bibinfo  {journal} {Electron Device
  Letters, IEEE}\ }\textbf {\bibinfo {volume} {23}},\ \bibinfo {pages} {386}
  (\bibinfo {year} {2002})}\BibitemShut {NoStop}%
\bibitem [{\citenamefont {Spanier}\ \emph {et~al.}(2006)\citenamefont
  {Spanier}, \citenamefont {Kolpak}, \citenamefont {Urban}, \citenamefont
  {Grinberg}, \citenamefont {Ouyang}, \citenamefont {Yun}, \citenamefont
  {Rappe},\ and\ \citenamefont {Park}}]{Spanier06p735}%
  \BibitemOpen
  \bibfield  {author} {\bibinfo {author} {\bibfnamefont {J.~E.}\ \bibnamefont
  {Spanier}}, \bibinfo {author} {\bibfnamefont {A.~M.}\ \bibnamefont {Kolpak}},
  \bibinfo {author} {\bibfnamefont {J.~J.}\ \bibnamefont {Urban}}, \bibinfo
  {author} {\bibfnamefont {I.}~\bibnamefont {Grinberg}}, \bibinfo {author}
  {\bibfnamefont {L.}~\bibnamefont {Ouyang}}, \bibinfo {author} {\bibfnamefont
  {W.~S.}\ \bibnamefont {Yun}}, \bibinfo {author} {\bibfnamefont {A.~M.}\
  \bibnamefont {Rappe}}, \ and\ \bibinfo {author} {\bibfnamefont
  {H.}~\bibnamefont {Park}},\ }\href@noop {} {\bibfield  {journal} {\bibinfo
  {journal} {Nano Lett.}\ }\textbf {\bibinfo {volume} {6}},\ \bibinfo {pages}
  {735} (\bibinfo {year} {2006})}\BibitemShut {NoStop}%
\bibitem [{\citenamefont {Urban}\ \emph {et~al.}(2002)\citenamefont {Urban},
  \citenamefont {Yun}, \citenamefont {Gu},\ and\ \citenamefont
  {Park}}]{Urban02p1186}%
  \BibitemOpen
  \bibfield  {author} {\bibinfo {author} {\bibfnamefont {J.~J.}\ \bibnamefont
  {Urban}}, \bibinfo {author} {\bibfnamefont {W.~S.}\ \bibnamefont {Yun}},
  \bibinfo {author} {\bibfnamefont {Q.}~\bibnamefont {Gu}}, \ and\ \bibinfo
  {author} {\bibfnamefont {H.}~\bibnamefont {Park}},\ }\href@noop {} {\bibfield
   {journal} {\bibinfo  {journal} {Journal of the American Chemical Society}\
  }\textbf {\bibinfo {volume} {124}},\ \bibinfo {pages} {1186} (\bibinfo {year}
  {2002})}\BibitemShut {NoStop}%
\bibitem [{\citenamefont {Yun}\ \emph {et~al.}(2002)\citenamefont {Yun},
  \citenamefont {Urban}, \citenamefont {Gu},\ and\ \citenamefont
  {Park}}]{Yun02p447}%
  \BibitemOpen
  \bibfield  {author} {\bibinfo {author} {\bibfnamefont {W.~S.}\ \bibnamefont
  {Yun}}, \bibinfo {author} {\bibfnamefont {J.~J.}\ \bibnamefont {Urban}},
  \bibinfo {author} {\bibfnamefont {Q.}~\bibnamefont {Gu}}, \ and\ \bibinfo
  {author} {\bibfnamefont {H.}~\bibnamefont {Park}},\ }\href@noop {} {\bibfield
   {journal} {\bibinfo  {journal} {Nano Lett.}\ }\textbf {\bibinfo {volume}
  {2}},\ \bibinfo {pages} {447} (\bibinfo {year} {2002})}\BibitemShut {NoStop}%
\bibitem [{\citenamefont {Urban}\ \emph {et~al.}(2003)\citenamefont {Urban},
  \citenamefont {Spanier}, \citenamefont {Ouyang}, \citenamefont {Yun},\ and\
  \citenamefont {Park}}]{Urban03p423}%
  \BibitemOpen
  \bibfield  {author} {\bibinfo {author} {\bibfnamefont {J.~J.}\ \bibnamefont
  {Urban}}, \bibinfo {author} {\bibfnamefont {J.~E.}\ \bibnamefont {Spanier}},
  \bibinfo {author} {\bibfnamefont {L.}~\bibnamefont {Ouyang}}, \bibinfo
  {author} {\bibfnamefont {W.~S.}\ \bibnamefont {Yun}}, \ and\ \bibinfo
  {author} {\bibfnamefont {H.}~\bibnamefont {Park}},\ }\href@noop {} {\bibfield
   {journal} {\bibinfo  {journal} {Adv. Mater.}\ }\textbf {\bibinfo {volume}
  {15}},\ \bibinfo {pages} {423} (\bibinfo {year} {2003})}\BibitemShut
  {NoStop}%
\bibitem [{\citenamefont {Batra}\ and\ \citenamefont
  {Silverman}(1972)}]{Batra72p291}%
  \BibitemOpen
  \bibfield  {author} {\bibinfo {author} {\bibfnamefont {I.~P.}\ \bibnamefont
  {Batra}}\ and\ \bibinfo {author} {\bibfnamefont {B.~D.}\ \bibnamefont
  {Silverman}},\ }\href@noop {} {\bibfield  {journal} {\bibinfo  {journal}
  {Solid State Commun.}\ }\textbf {\bibinfo {volume} {11}},\ \bibinfo {pages}
  {291} (\bibinfo {year} {1972})}\BibitemShut {NoStop}%
\bibitem [{\citenamefont {Nonnenmann}\ and\ \citenamefont
  {Spanier}(2009)}]{Nonnenmann09p5205}%
  \BibitemOpen
  \bibfield  {author} {\bibinfo {author} {\bibfnamefont {S.~S.}\ \bibnamefont
  {Nonnenmann}}\ and\ \bibinfo {author} {\bibfnamefont {J.~E.}\ \bibnamefont
  {Spanier}},\ }\href@noop {} {\bibfield  {journal} {\bibinfo  {journal}
  {Journal of materials science}\ }\textbf {\bibinfo {volume} {44}},\ \bibinfo
  {pages} {5205} (\bibinfo {year} {2009})}\BibitemShut {NoStop}%
\bibitem [{\citenamefont {Stengel}\ and\ \citenamefont
  {Spaldin}(2006)}]{Stengel06p679}%
  \BibitemOpen
  \bibfield  {author} {\bibinfo {author} {\bibfnamefont {M.}~\bibnamefont
  {Stengel}}\ and\ \bibinfo {author} {\bibfnamefont {N.}~\bibnamefont
  {Spaldin}},\ }\href@noop {} {\bibfield  {journal} {\bibinfo  {journal}
  {Nature}\ }\textbf {\bibinfo {volume} {443}},\ \bibinfo {pages} {679}
  (\bibinfo {year} {2006})}\BibitemShut {NoStop}%
\bibitem [{\citenamefont {Szwarcman}\ \emph {et~al.}(2012)\citenamefont
  {Szwarcman}, \citenamefont {Lubk}, \citenamefont {Linck}, \citenamefont
  {Vogel}, \citenamefont {Lereah}, \citenamefont {Lichte},\ and\ \citenamefont
  {Markovich}}]{Szwarcman12p134112}%
  \BibitemOpen
  \bibfield  {author} {\bibinfo {author} {\bibfnamefont {D.}~\bibnamefont
  {Szwarcman}}, \bibinfo {author} {\bibfnamefont {A.}~\bibnamefont {Lubk}},
  \bibinfo {author} {\bibfnamefont {M.}~\bibnamefont {Linck}}, \bibinfo
  {author} {\bibfnamefont {K.}~\bibnamefont {Vogel}}, \bibinfo {author}
  {\bibfnamefont {Y.}~\bibnamefont {Lereah}}, \bibinfo {author} {\bibfnamefont
  {H.}~\bibnamefont {Lichte}}, \ and\ \bibinfo {author} {\bibfnamefont
  {G.}~\bibnamefont {Markovich}},\ }\href@noop {} {\bibfield  {journal}
  {\bibinfo  {journal} {Physical Review B}\ }\textbf {\bibinfo {volume} {85}},\
  \bibinfo {pages} {134112} (\bibinfo {year} {2012})}\BibitemShut {NoStop}%
\bibitem [{\citenamefont {Szwarcman}\ \emph {et~al.}(2014)\citenamefont
  {Szwarcman}, \citenamefont {Prosandeev}, \citenamefont {Louis}, \citenamefont
  {Berger}, \citenamefont {Rosenberg}, \citenamefont {Lereah}, \citenamefont
  {Bellaiche},\ and\ \citenamefont {Markovich}}]{Szwarcman14p122202}%
  \BibitemOpen
  \bibfield  {author} {\bibinfo {author} {\bibfnamefont {D.}~\bibnamefont
  {Szwarcman}}, \bibinfo {author} {\bibfnamefont {S.}~\bibnamefont
  {Prosandeev}}, \bibinfo {author} {\bibfnamefont {L.}~\bibnamefont {Louis}},
  \bibinfo {author} {\bibfnamefont {S.}~\bibnamefont {Berger}}, \bibinfo
  {author} {\bibfnamefont {Y.}~\bibnamefont {Rosenberg}}, \bibinfo {author}
  {\bibfnamefont {Y.}~\bibnamefont {Lereah}}, \bibinfo {author} {\bibfnamefont
  {L.}~\bibnamefont {Bellaiche}}, \ and\ \bibinfo {author} {\bibfnamefont
  {G.}~\bibnamefont {Markovich}},\ }\href@noop {} {\bibfield  {journal}
  {\bibinfo  {journal} {Journal of Physics: Condensed Matter}\ }\textbf
  {\bibinfo {volume} {26}},\ \bibinfo {pages} {122202} (\bibinfo {year}
  {2014})}\BibitemShut {NoStop}%
\bibitem [{\citenamefont {Kim}\ \emph {et~al.}(2005)\citenamefont {Kim},
  \citenamefont {Jo}, \citenamefont {Kim}, \citenamefont {Chang}, \citenamefont
  {Lee}, \citenamefont {Yoon}, \citenamefont {Song},\ and\ \citenamefont
  {Noh}}]{Kim05p237602}%
  \BibitemOpen
  \bibfield  {author} {\bibinfo {author} {\bibfnamefont {D.}~\bibnamefont
  {Kim}}, \bibinfo {author} {\bibfnamefont {J.}~\bibnamefont {Jo}}, \bibinfo
  {author} {\bibfnamefont {Y.}~\bibnamefont {Kim}}, \bibinfo {author}
  {\bibfnamefont {Y.}~\bibnamefont {Chang}}, \bibinfo {author} {\bibfnamefont
  {J.}~\bibnamefont {Lee}}, \bibinfo {author} {\bibfnamefont {J.-G.}\
  \bibnamefont {Yoon}}, \bibinfo {author} {\bibfnamefont {T.}~\bibnamefont
  {Song}}, \ and\ \bibinfo {author} {\bibfnamefont {T.}~\bibnamefont {Noh}},\
  }\href@noop {} {\bibfield  {journal} {\bibinfo  {journal} {Physical review
  letters}\ }\textbf {\bibinfo {volume} {95}},\ \bibinfo {pages} {237602}
  (\bibinfo {year} {2005})}\BibitemShut {NoStop}%
\bibitem [{\citenamefont {Saidi}\ \emph {et~al.}(2014)\citenamefont {Saidi},
  \citenamefont {Martirez},\ and\ \citenamefont {Rappe}}]{Saidi14p6711}%
  \BibitemOpen
  \bibfield  {author} {\bibinfo {author} {\bibfnamefont {W.~A.}\ \bibnamefont
  {Saidi}}, \bibinfo {author} {\bibfnamefont {J.~M.~P.}\ \bibnamefont
  {Martirez}}, \ and\ \bibinfo {author} {\bibfnamefont {A.~M.}\ \bibnamefont
  {Rappe}},\ }\href@noop {} {\bibfield  {journal} {\bibinfo  {journal} {Nano
  letters}\ }\textbf {\bibinfo {volume} {14}},\ \bibinfo {pages} {6711}
  (\bibinfo {year} {2014})}\BibitemShut {NoStop}%
\bibitem [{\citenamefont {Kolpak}\ \emph {et~al.}(2008)\citenamefont {Kolpak},
  \citenamefont {Li}, \citenamefont {Shao}, \citenamefont {Rappe},\ and\
  \citenamefont {Bonnell}}]{Kolpak08p036102}%
  \BibitemOpen
  \bibfield  {author} {\bibinfo {author} {\bibfnamefont {A.~M.}\ \bibnamefont
  {Kolpak}}, \bibinfo {author} {\bibfnamefont {D.}~\bibnamefont {Li}}, \bibinfo
  {author} {\bibfnamefont {R.}~\bibnamefont {Shao}}, \bibinfo {author}
  {\bibfnamefont {A.~M.}\ \bibnamefont {Rappe}}, \ and\ \bibinfo {author}
  {\bibfnamefont {D.~A.}\ \bibnamefont {Bonnell}},\ }\href@noop {} {\bibfield
  {journal} {\bibinfo  {journal} {Phys. Rev. Lett.}\ }\textbf {\bibinfo
  {volume} {101}},\ \bibinfo {pages} {036102} (\bibinfo {year}
  {2008})}\BibitemShut {NoStop}%
\bibitem [{\citenamefont {Mendez-Polanco}\ \emph {et~al.}(2012)\citenamefont
  {Mendez-Polanco}, \citenamefont {Grinberg}, \citenamefont {Kolpak},
  \citenamefont {Levchenko}, \citenamefont {Pynn},\ and\ \citenamefont
  {Rappe}}]{Mendez-Polanco12p214107}%
  \BibitemOpen
  \bibfield  {author} {\bibinfo {author} {\bibfnamefont {M.~A.}\ \bibnamefont
  {Mendez-Polanco}}, \bibinfo {author} {\bibfnamefont {I.}~\bibnamefont
  {Grinberg}}, \bibinfo {author} {\bibfnamefont {A.~M.}\ \bibnamefont
  {Kolpak}}, \bibinfo {author} {\bibfnamefont {S.~V.}\ \bibnamefont
  {Levchenko}}, \bibinfo {author} {\bibfnamefont {C.}~\bibnamefont {Pynn}}, \
  and\ \bibinfo {author} {\bibfnamefont {A.~M.}\ \bibnamefont {Rappe}},\
  }\href@noop {} {\bibfield  {journal} {\bibinfo  {journal} {Phys. Rev. B}\
  }\textbf {\bibinfo {volume} {85}},\ \bibinfo {pages} {214107} (\bibinfo
  {year} {2012})}\BibitemShut {NoStop}%
\bibitem [{\citenamefont {Sai}\ \emph {et~al.}(2005)\citenamefont {Sai},
  \citenamefont {Kolpak},\ and\ \citenamefont {Rappe}}]{Sai05p020101R}%
  \BibitemOpen
  \bibfield  {author} {\bibinfo {author} {\bibfnamefont {N.}~\bibnamefont
  {Sai}}, \bibinfo {author} {\bibfnamefont {A.~M.}\ \bibnamefont {Kolpak}}, \
  and\ \bibinfo {author} {\bibfnamefont {A.~M.}\ \bibnamefont {Rappe}},\
  }\href@noop {} {\bibfield  {journal} {\bibinfo  {journal} {Phys. Rev. B Rapid
  Comm.}\ }\textbf {\bibinfo {volume} {72}},\ \bibinfo {pages} {020101(R)}
  (\bibinfo {year} {2005})}\BibitemShut {NoStop}%
\bibitem [{\citenamefont {Kolpak}\ \emph {et~al.}(2006)\citenamefont {Kolpak},
  \citenamefont {Sai},\ and\ \citenamefont {Rappe}}]{Kolpak06p054112}%
  \BibitemOpen
  \bibfield  {author} {\bibinfo {author} {\bibfnamefont {A.~M.}\ \bibnamefont
  {Kolpak}}, \bibinfo {author} {\bibfnamefont {N.}~\bibnamefont {Sai}}, \ and\
  \bibinfo {author} {\bibfnamefont {A.~M.}\ \bibnamefont {Rappe}},\ }\href@noop
  {} {\bibfield  {journal} {\bibinfo  {journal} {Phys. Rev. B}\ }\textbf
  {\bibinfo {volume} {74}},\ \bibinfo {pages} {054112} (\bibinfo {year}
  {2006})}\BibitemShut {NoStop}%
\bibitem [{\citenamefont {Stengel}\ \emph {et~al.}(2009)\citenamefont
  {Stengel}, \citenamefont {Vanderbilt},\ and\ \citenamefont
  {Spaldin}}]{Stengel09p392}%
  \BibitemOpen
  \bibfield  {author} {\bibinfo {author} {\bibfnamefont {M.}~\bibnamefont
  {Stengel}}, \bibinfo {author} {\bibfnamefont {D.}~\bibnamefont {Vanderbilt}},
  \ and\ \bibinfo {author} {\bibfnamefont {N.}~\bibnamefont {Spaldin}},\
  }\href@noop {} {\bibfield  {journal} {\bibinfo  {journal} {Nat. Mater.}\
  }\textbf {\bibinfo {volume} {8}},\ \bibinfo {pages} {392} (\bibinfo {year}
  {2009})}\BibitemShut {NoStop}%
\bibitem [{\citenamefont {Fong}\ \emph {et~al.}(2006)\citenamefont {Fong},
  \citenamefont {Kolpak}, \citenamefont {Eastman}, \citenamefont {Streiffer},
  \citenamefont {Fuoss}, \citenamefont {Stephenson}, \citenamefont {Thompson},
  \citenamefont {Kim}, \citenamefont {Choi}, \citenamefont {Eom}, \citenamefont
  {Grinberg},\ and\ \citenamefont {Rappe}}]{Fong06p127601}%
  \BibitemOpen
  \bibfield  {author} {\bibinfo {author} {\bibfnamefont {D.~D.}\ \bibnamefont
  {Fong}}, \bibinfo {author} {\bibfnamefont {A.~M.}\ \bibnamefont {Kolpak}},
  \bibinfo {author} {\bibfnamefont {J.~A.}\ \bibnamefont {Eastman}}, \bibinfo
  {author} {\bibfnamefont {S.~K.}\ \bibnamefont {Streiffer}}, \bibinfo {author}
  {\bibfnamefont {P.~H.}\ \bibnamefont {Fuoss}}, \bibinfo {author}
  {\bibfnamefont {G.~B.}\ \bibnamefont {Stephenson}}, \bibinfo {author}
  {\bibfnamefont {C.}~\bibnamefont {Thompson}}, \bibinfo {author}
  {\bibfnamefont {D.~M.}\ \bibnamefont {Kim}}, \bibinfo {author} {\bibfnamefont
  {K.~J.}\ \bibnamefont {Choi}}, \bibinfo {author} {\bibfnamefont {C.~B.}\
  \bibnamefont {Eom}}, \bibinfo {author} {\bibfnamefont {I.}~\bibnamefont
  {Grinberg}}, \ and\ \bibinfo {author} {\bibfnamefont {A.~M.}\ \bibnamefont
  {Rappe}},\ }\href@noop {} {\bibfield  {journal} {\bibinfo  {journal} {Phys.
  Rev. Lett.}\ }\textbf {\bibinfo {volume} {96}},\ \bibinfo {pages} {127601}
  (\bibinfo {year} {2006})}\BibitemShut {NoStop}%
\bibitem [{\citenamefont {Li}\ \emph {et~al.}(2008)\citenamefont {Li},
  \citenamefont {Zhao}, \citenamefont {Garra}, \citenamefont {Kolpak},
  \citenamefont {Rappe}, \citenamefont {Bonnell},\ and\ \citenamefont
  {Vohs}}]{Li08p473}%
  \BibitemOpen
  \bibfield  {author} {\bibinfo {author} {\bibfnamefont {D.}~\bibnamefont
  {Li}}, \bibinfo {author} {\bibfnamefont {M.~H.}\ \bibnamefont {Zhao}},
  \bibinfo {author} {\bibfnamefont {J.}~\bibnamefont {Garra}}, \bibinfo
  {author} {\bibfnamefont {A.}~\bibnamefont {Kolpak}}, \bibinfo {author}
  {\bibfnamefont {A.}~\bibnamefont {Rappe}}, \bibinfo {author} {\bibfnamefont
  {D.~A.}\ \bibnamefont {Bonnell}}, \ and\ \bibinfo {author} {\bibfnamefont
  {J.~M.}\ \bibnamefont {Vohs}},\ }\href@noop {} {\bibfield  {journal}
  {\bibinfo  {journal} {Nature Mater.}\ }\textbf {\bibinfo {volume} {7}},\
  \bibinfo {pages} {473} (\bibinfo {year} {2008})}\BibitemShut {NoStop}%
\bibitem [{\citenamefont {Wang}\ \emph {et~al.}(2009)\citenamefont {Wang},
  \citenamefont {Fong}, \citenamefont {Jiang}, \citenamefont {Highland},
  \citenamefont {Fuoss}, \citenamefont {Thompson}, \citenamefont {Kolpak},
  \citenamefont {Eastman}, \citenamefont {Streiffer}, \citenamefont {Rappe},\
  and\ \citenamefont {Stephenson}}]{Wang09p047601}%
  \BibitemOpen
  \bibfield  {author} {\bibinfo {author} {\bibfnamefont {R.~V.}\ \bibnamefont
  {Wang}}, \bibinfo {author} {\bibfnamefont {D.~D.}\ \bibnamefont {Fong}},
  \bibinfo {author} {\bibfnamefont {F.}~\bibnamefont {Jiang}}, \bibinfo
  {author} {\bibfnamefont {M.~J.}\ \bibnamefont {Highland}}, \bibinfo {author}
  {\bibfnamefont {P.~H.}\ \bibnamefont {Fuoss}}, \bibinfo {author}
  {\bibfnamefont {C.}~\bibnamefont {Thompson}}, \bibinfo {author}
  {\bibfnamefont {A.~M.}\ \bibnamefont {Kolpak}}, \bibinfo {author}
  {\bibfnamefont {J.~A.}\ \bibnamefont {Eastman}}, \bibinfo {author}
  {\bibfnamefont {S.~K.}\ \bibnamefont {Streiffer}}, \bibinfo {author}
  {\bibfnamefont {A.~M.}\ \bibnamefont {Rappe}}, \ and\ \bibinfo {author}
  {\bibfnamefont {G.~B.}\ \bibnamefont {Stephenson}},\ }\href@noop {}
  {\bibfield  {journal} {\bibinfo  {journal} {Phys. Rev. Lett..}\ }\textbf
  {\bibinfo {volume} {102}},\ \bibinfo {pages} {047601 1} (\bibinfo {year}
  {2009})}\BibitemShut {NoStop}%
\bibitem [{\citenamefont {Koocher}\ \emph {et~al.}(2014)\citenamefont
  {Koocher}, \citenamefont {Martirez},\ and\ \citenamefont
  {Rappe}}]{Koocher14p3408}%
  \BibitemOpen
  \bibfield  {author} {\bibinfo {author} {\bibfnamefont {N.~Z.}\ \bibnamefont
  {Koocher}}, \bibinfo {author} {\bibfnamefont {J.~M.~P.}\ \bibnamefont
  {Martirez}}, \ and\ \bibinfo {author} {\bibfnamefont {A.~M.}\ \bibnamefont
  {Rappe}},\ }\href@noop {} {\bibfield  {journal} {\bibinfo  {journal} {The
  Journal of Physical Chemistry Letters}\ }\textbf {\bibinfo {volume} {5}},\
  \bibinfo {pages} {3408} (\bibinfo {year} {2014})}\BibitemShut {NoStop}%
\bibitem [{\citenamefont {Stephenson}\ and\ \citenamefont
  {Highland}(2011)}]{Stephenson11p064107}%
  \BibitemOpen
  \bibfield  {author} {\bibinfo {author} {\bibfnamefont {G.~B.}\ \bibnamefont
  {Stephenson}}\ and\ \bibinfo {author} {\bibfnamefont {M.~J.}\ \bibnamefont
  {Highland}},\ }\href@noop {} {\bibfield  {journal} {\bibinfo  {journal}
  {Physical Review B}\ }\textbf {\bibinfo {volume} {84}},\ \bibinfo {pages}
  {064107} (\bibinfo {year} {2011})}\BibitemShut {NoStop}%
\bibitem [{\citenamefont {Levchenko}\ and\ \citenamefont
  {Rappe}(2008)}]{Levchenko08p256101}%
  \BibitemOpen
  \bibfield  {author} {\bibinfo {author} {\bibfnamefont {S.~V.}\ \bibnamefont
  {Levchenko}}\ and\ \bibinfo {author} {\bibfnamefont {A.~M.}\ \bibnamefont
  {Rappe}},\ }\href@noop {} {\bibfield  {journal} {\bibinfo  {journal} {Phys.
  Rev. Lett.}\ }\textbf {\bibinfo {volume} {100}},\ \bibinfo {pages} {256101}
  (\bibinfo {year} {2008})}\BibitemShut {NoStop}%
\bibitem [{\citenamefont {He}\ \emph {et~al.}(2011{\natexlab{a}})\citenamefont
  {He}, \citenamefont {Qiao}, \citenamefont {Volinsky}, \citenamefont {Bai},
  \citenamefont {Wu},\ and\ \citenamefont {Chu}}]{He11p062905}%
  \BibitemOpen
  \bibfield  {author} {\bibinfo {author} {\bibfnamefont {D.}~\bibnamefont
  {He}}, \bibinfo {author} {\bibfnamefont {L.}~\bibnamefont {Qiao}}, \bibinfo
  {author} {\bibfnamefont {A.~A.}\ \bibnamefont {Volinsky}}, \bibinfo {author}
  {\bibfnamefont {Y.}~\bibnamefont {Bai}}, \bibinfo {author} {\bibfnamefont
  {M.}~\bibnamefont {Wu}}, \ and\ \bibinfo {author} {\bibfnamefont
  {W.}~\bibnamefont {Chu}},\ }\href@noop {} {\bibfield  {journal} {\bibinfo
  {journal} {Applied Physics Letters}\ }\textbf {\bibinfo {volume} {98}},\
  \bibinfo {pages} {062905} (\bibinfo {year} {2011}{\natexlab{a}})}\BibitemShut
  {NoStop}%
\bibitem [{\citenamefont {He}\ \emph {et~al.}(2011{\natexlab{b}})\citenamefont
  {He}, \citenamefont {Qiao}, \citenamefont {Volinsky}, \citenamefont {Bai},\
  and\ \citenamefont {Guo}}]{He11p024101}%
  \BibitemOpen
  \bibfield  {author} {\bibinfo {author} {\bibfnamefont {D.}~\bibnamefont
  {He}}, \bibinfo {author} {\bibfnamefont {L.}~\bibnamefont {Qiao}}, \bibinfo
  {author} {\bibfnamefont {A.~A.}\ \bibnamefont {Volinsky}}, \bibinfo {author}
  {\bibfnamefont {Y.}~\bibnamefont {Bai}}, \ and\ \bibinfo {author}
  {\bibfnamefont {L.}~\bibnamefont {Guo}},\ }\href@noop {} {\bibfield
  {journal} {\bibinfo  {journal} {Physical Review B}\ }\textbf {\bibinfo
  {volume} {84}},\ \bibinfo {pages} {024101} (\bibinfo {year}
  {2011}{\natexlab{b}})}\BibitemShut {NoStop}%
\bibitem [{\citenamefont {Kalinin}\ \emph {et~al.}(2002)\citenamefont
  {Kalinin}, \citenamefont {Johnson},\ and\ \citenamefont
  {Bonnell}}]{Kalinin02p3816}%
  \BibitemOpen
  \bibfield  {author} {\bibinfo {author} {\bibfnamefont {S.~V.}\ \bibnamefont
  {Kalinin}}, \bibinfo {author} {\bibfnamefont {C.~Y.}\ \bibnamefont
  {Johnson}}, \ and\ \bibinfo {author} {\bibfnamefont {D.~A.}\ \bibnamefont
  {Bonnell}},\ }\href@noop {} {\bibfield  {journal} {\bibinfo  {journal} {J.
  Appl. Phys.}\ }\textbf {\bibinfo {volume} {91}},\ \bibinfo {pages} {3816}
  (\bibinfo {year} {2002})}\BibitemShut {NoStop}%
\bibitem [{\citenamefont {Noma}\ \emph {et~al.}(1996)\citenamefont {Noma},
  \citenamefont {Wada}, \citenamefont {Yano},\ and\ \citenamefont
  {Suzuki}}]{Noma96p5223}%
  \BibitemOpen
  \bibfield  {author} {\bibinfo {author} {\bibfnamefont {T.}~\bibnamefont
  {Noma}}, \bibinfo {author} {\bibfnamefont {S.}~\bibnamefont {Wada}}, \bibinfo
  {author} {\bibfnamefont {M.}~\bibnamefont {Yano}}, \ and\ \bibinfo {author}
  {\bibfnamefont {T.}~\bibnamefont {Suzuki}},\ }\href@noop {} {\bibfield
  {journal} {\bibinfo  {journal} {J. Appl. Phys.}\ }\textbf {\bibinfo {volume}
  {80}},\ \bibinfo {pages} {5223} (\bibinfo {year} {1996})}\BibitemShut
  {NoStop}%
\bibitem [{\citenamefont {Wegmann}\ \emph {et~al.}(2004)\citenamefont
  {Wegmann}, \citenamefont {Watson},\ and\ \citenamefont
  {Hendry}}]{Wegmann04p371}%
  \BibitemOpen
  \bibfield  {author} {\bibinfo {author} {\bibfnamefont {M.}~\bibnamefont
  {Wegmann}}, \bibinfo {author} {\bibfnamefont {L.}~\bibnamefont {Watson}}, \
  and\ \bibinfo {author} {\bibfnamefont {A.}~\bibnamefont {Hendry}},\
  }\href@noop {} {\bibfield  {journal} {\bibinfo  {journal} {Journal of the
  American Ceramic Society}\ }\textbf {\bibinfo {volume} {87}},\ \bibinfo
  {pages} {371} (\bibinfo {year} {2004})}\BibitemShut {NoStop}%
\bibitem [{Opi()}]{Opium}%
  \BibitemOpen
  \href@noop {} {}\bibinfo {howpublished}
  {http://opium.sourceforge.net}\BibitemShut {NoStop}%
\bibitem [{\citenamefont {Al-Saidi}\ and\ \citenamefont
  {Rappe}(2010)}]{Al-Saidi10p155304}%
  \BibitemOpen
  \bibfield  {author} {\bibinfo {author} {\bibfnamefont {W.~A.}\ \bibnamefont
  {Al-Saidi}}\ and\ \bibinfo {author} {\bibfnamefont {A.~M.}\ \bibnamefont
  {Rappe}},\ }\href {\doibase 10.1103/PhysRevB.82.155304} {\bibfield  {journal}
  {\bibinfo  {journal} {Phys. Rev. B}\ }\textbf {\bibinfo {volume} {82}},\
  \bibinfo {pages} {155304} (\bibinfo {year} {2010})}\BibitemShut {NoStop}%
\bibitem [{\citenamefont {Burstein}\ and\ \citenamefont
  {Lundqvist}(1969)}]{Burstein1969tunneling}%
  \BibitemOpen
  \bibfield  {author} {\bibinfo {author} {\bibfnamefont {E.}~\bibnamefont
  {Burstein}}\ and\ \bibinfo {author} {\bibfnamefont {S.}~\bibnamefont
  {Lundqvist}},\ }\href@noop {} {\emph {\bibinfo {title} {Tunneling phenomena
  in solids}}}\ (\bibinfo  {publisher} {Plenum Press},\ \bibinfo {address} {New
  York},\ \bibinfo {year} {1969})\BibitemShut {NoStop}%
\bibitem [{\citenamefont {Dekker}(1957)}]{Dekkersolid}%
  \BibitemOpen
  \bibfield  {author} {\bibinfo {author} {\bibfnamefont {A.}~\bibnamefont
  {Dekker}},\ }\href@noop {} {\emph {\bibinfo {title} {Solid State Physics}}}\
  (\bibinfo  {publisher} {Prentice--Hall, INC.},\ \bibinfo {address} {Englewood
  Cliffs, N. J.},\ \bibinfo {year} {1957})\BibitemShut {NoStop}%
\bibitem [{\citenamefont {Simmons}(1965)}]{Simmons65p967}%
  \BibitemOpen
  \bibfield  {author} {\bibinfo {author} {\bibfnamefont {J.}~\bibnamefont
  {Simmons}},\ }\href@noop {} {\bibfield  {journal} {\bibinfo  {journal}
  {Physical Review Letters}\ }\textbf {\bibinfo {volume} {15}},\ \bibinfo
  {pages} {967} (\bibinfo {year} {1965})}\BibitemShut {NoStop}%
\bibitem [{\citenamefont {Pyykko}\ and\ \citenamefont
  {Laaksonen}(1984)}]{Pyykko84p4892}%
  \BibitemOpen
  \bibfield  {author} {\bibinfo {author} {\bibfnamefont {P.}~\bibnamefont
  {Pyykko}}\ and\ \bibinfo {author} {\bibfnamefont {L.}~\bibnamefont
  {Laaksonen}},\ }\href@noop {} {\bibfield  {journal} {\bibinfo  {journal} {The
  Journal of Physical Chemistry}\ }\textbf {\bibinfo {volume} {88}},\ \bibinfo
  {pages} {4892} (\bibinfo {year} {1984})}\BibitemShut {NoStop}%
\bibitem [{\citenamefont {Rappe}\ \emph {et~al.}(1990)\citenamefont {Rappe},
  \citenamefont {Rabe}, \citenamefont {Kaxiras},\ and\ \citenamefont
  {Joannopoulos}}]{Rappe90p1227}%
  \BibitemOpen
  \bibfield  {author} {\bibinfo {author} {\bibfnamefont {A.~M.}\ \bibnamefont
  {Rappe}}, \bibinfo {author} {\bibfnamefont {K.~M.}\ \bibnamefont {Rabe}},
  \bibinfo {author} {\bibfnamefont {E.}~\bibnamefont {Kaxiras}}, \ and\
  \bibinfo {author} {\bibfnamefont {J.~D.}\ \bibnamefont {Joannopoulos}},\
  }\href@noop {} {\bibfield  {journal} {\bibinfo  {journal} {Phys. Rev. B Rapid
  Comm.}\ }\textbf {\bibinfo {volume} {41}},\ \bibinfo {pages} {1227} (\bibinfo
  {year} {1990})}\BibitemShut {NoStop}%
\bibitem [{\citenamefont {McQuarrie}\ and\ \citenamefont
  {Simon}(1997)}]{Mcquarrie1997physical}%
  \BibitemOpen
  \bibfield  {author} {\bibinfo {author} {\bibfnamefont {D.~D.~A.}\
  \bibnamefont {McQuarrie}}\ and\ \bibinfo {author} {\bibfnamefont {J.~J.~D.}\
  \bibnamefont {Simon}},\ }\href@noop {} {\emph {\bibinfo {title} {Physical
  chemistry: a molecular approach}}}\ (\bibinfo  {publisher} {University
  Science Books},\ \bibinfo {year} {1997})\BibitemShut {NoStop}%
\bibitem [{\citenamefont {Zafar}\ \emph {et~al.}(1998)\citenamefont {Zafar},
  \citenamefont {Jones}, \citenamefont {Jiang}, \citenamefont {White},
  \citenamefont {Kaushik},\ and\ \citenamefont {Gillespie}}]{Zafar98p3533}%
  \BibitemOpen
  \bibfield  {author} {\bibinfo {author} {\bibfnamefont {S.}~\bibnamefont
  {Zafar}}, \bibinfo {author} {\bibfnamefont {R.~E.}\ \bibnamefont {Jones}},
  \bibinfo {author} {\bibfnamefont {B.}~\bibnamefont {Jiang}}, \bibinfo
  {author} {\bibfnamefont {B.}~\bibnamefont {White}}, \bibinfo {author}
  {\bibfnamefont {V.}~\bibnamefont {Kaushik}}, \ and\ \bibinfo {author}
  {\bibfnamefont {S.}~\bibnamefont {Gillespie}},\ }\href@noop {} {\bibfield
  {journal} {\bibinfo  {journal} {Applied physics letters}\ }\textbf {\bibinfo
  {volume} {73}},\ \bibinfo {pages} {3533} (\bibinfo {year}
  {1998})}\BibitemShut {NoStop}%
\bibitem [{\citenamefont {Ziman}(1960)}]{Ziman1960electronics}%
  \BibitemOpen
  \bibfield  {author} {\bibinfo {author} {\bibfnamefont {J.}~\bibnamefont
  {Ziman}},\ }\href@noop {} {\emph {\bibinfo {title} {Electronics and
  Phonons}}}\ (\bibinfo  {publisher} {Clarendon Press, Oxford},\ \bibinfo
  {address} {London},\ \bibinfo {year} {1960})\BibitemShut {NoStop}%
\bibitem [{\citenamefont {Low}\ and\ \citenamefont {Pines}(1955)}]{Low55p414}%
  \BibitemOpen
  \bibfield  {author} {\bibinfo {author} {\bibfnamefont {F.~E.}\ \bibnamefont
  {Low}}\ and\ \bibinfo {author} {\bibfnamefont {D.}~\bibnamefont {Pines}},\
  }\href@noop {} {\bibfield  {journal} {\bibinfo  {journal} {Physical Review}\
  }\textbf {\bibinfo {volume} {98}},\ \bibinfo {pages} {414} (\bibinfo {year}
  {1955})}\BibitemShut {NoStop}%
\bibitem [{\citenamefont {Schultz}(1959)}]{Schultz59p526}%
  \BibitemOpen
  \bibfield  {author} {\bibinfo {author} {\bibfnamefont {T.}~\bibnamefont
  {Schultz}},\ }\href@noop {} {\bibfield  {journal} {\bibinfo  {journal}
  {Physical Review}\ }\textbf {\bibinfo {volume} {116}},\ \bibinfo {pages}
  {526} (\bibinfo {year} {1959})}\BibitemShut {NoStop}%
\bibitem [{\citenamefont {Wemple}\ \emph {et~al.}(1969)\citenamefont {Wemple},
  \citenamefont {DiDomenico~Jr},\ and\ \citenamefont
  {Jayaraman}}]{Wemple69p547}%
  \BibitemOpen
  \bibfield  {author} {\bibinfo {author} {\bibfnamefont {S.}~\bibnamefont
  {Wemple}}, \bibinfo {author} {\bibfnamefont {M.}~\bibnamefont
  {DiDomenico~Jr}}, \ and\ \bibinfo {author} {\bibfnamefont {A.}~\bibnamefont
  {Jayaraman}},\ }\href@noop {} {\bibfield  {journal} {\bibinfo  {journal}
  {Physical Review}\ }\textbf {\bibinfo {volume} {180}},\ \bibinfo {pages}
  {547} (\bibinfo {year} {1969})}\BibitemShut {NoStop}%
\bibitem [{\citenamefont {Spigler}\ and\ \citenamefont
  {Vianello}(1995)}]{Spigler95p567}%
  \BibitemOpen
  \bibfield  {author} {\bibinfo {author} {\bibfnamefont {R.}~\bibnamefont
  {Spigler}}\ and\ \bibinfo {author} {\bibfnamefont {M.}~\bibnamefont
  {Vianello}},\ }\href@noop {} {\bibfield  {journal} {\bibinfo  {journal}
  {Advances in Difference Equations (Veszpr{\'e}m, 1995)}\ ,\ \bibinfo {pages}
  {567}} (\bibinfo {year} {1995})}\BibitemShut {NoStop}%
\bibitem [{\citenamefont {Lyddane}\ \emph {et~al.}(1941)\citenamefont
  {Lyddane}, \citenamefont {Sachs},\ and\ \citenamefont
  {Teller}}]{Lyddane41p673}%
  \BibitemOpen
  \bibfield  {author} {\bibinfo {author} {\bibfnamefont {R.~H.}\ \bibnamefont
  {Lyddane}}, \bibinfo {author} {\bibfnamefont {R.}~\bibnamefont {Sachs}}, \
  and\ \bibinfo {author} {\bibfnamefont {E.}~\bibnamefont {Teller}},\
  }\href@noop {} {\bibfield  {journal} {\bibinfo  {journal} {Phys. Rev.}\
  }\textbf {\bibinfo {volume} {59}},\ \bibinfo {pages} {673} (\bibinfo {year}
  {1941})}\BibitemShut {NoStop}%
\bibitem [{\citenamefont {Shirane}\ \emph {et~al.}(1967)\citenamefont
  {Shirane}, \citenamefont {Frazer}, \citenamefont {Minkiewicz}, \citenamefont
  {Leake},\ and\ \citenamefont {Linz}}]{Shirane67p234}%
  \BibitemOpen
  \bibfield  {author} {\bibinfo {author} {\bibfnamefont {G.}~\bibnamefont
  {Shirane}}, \bibinfo {author} {\bibfnamefont {B.}~\bibnamefont {Frazer}},
  \bibinfo {author} {\bibfnamefont {V.}~\bibnamefont {Minkiewicz}}, \bibinfo
  {author} {\bibfnamefont {J.}~\bibnamefont {Leake}}, \ and\ \bibinfo {author}
  {\bibfnamefont {A.}~\bibnamefont {Linz}},\ }\href@noop {} {\bibfield
  {journal} {\bibinfo  {journal} {Physical Review Letters}\ }\textbf {\bibinfo
  {volume} {19}},\ \bibinfo {pages} {234} (\bibinfo {year} {1967})}\BibitemShut
  {NoStop}%
\bibitem [{\citenamefont {Rabe}\ \emph {et~al.}(2007)\citenamefont {Rabe},
  \citenamefont {Ahn},\ and\ \citenamefont {Triscone}}]{Rabe07}%
  \BibitemOpen
  \bibfield  {author} {\bibinfo {author} {\bibfnamefont {K.~M.}\ \bibnamefont
  {Rabe}}, \bibinfo {author} {\bibfnamefont {C.~H.}\ \bibnamefont {Ahn}}, \
  and\ \bibinfo {author} {\bibfnamefont {J.-M.}\ \bibnamefont {Triscone}},\
  }\href@noop {} {\emph {\bibinfo {title} {Physics of Ferroelectrics: A Modern
  Perspective}}}\ (\bibinfo  {publisher} {Springer-Verlag},\ \bibinfo {year}
  {2007})\BibitemShut {NoStop}%
\bibitem [{\citenamefont {Singh}(1996)}]{Singh96p176}%
  \BibitemOpen
  \bibfield  {author} {\bibinfo {author} {\bibfnamefont {D.~J.}\ \bibnamefont
  {Singh}},\ }\href@noop {} {\bibfield  {journal} {\bibinfo  {journal} {Phys.
  Rev. B}\ }\textbf {\bibinfo {volume} {53}},\ \bibinfo {pages} {176} (\bibinfo
  {year} {1996})}\BibitemShut {NoStop}%
\bibitem [{\citenamefont {Benedict}\ and\ \citenamefont
  {Durand}(1958)}]{Benedict59p1091}%
  \BibitemOpen
  \bibfield  {author} {\bibinfo {author} {\bibfnamefont {T.}~\bibnamefont
  {Benedict}}\ and\ \bibinfo {author} {\bibfnamefont {J.}~\bibnamefont
  {Durand}},\ }\href@noop {} {\bibfield  {journal} {\bibinfo  {journal}
  {Physical Review}\ }\textbf {\bibinfo {volume} {109}},\ \bibinfo {pages}
  {1091} (\bibinfo {year} {1958})}\BibitemShut {NoStop}%
\bibitem [{\citenamefont {Barker~Jr}(1966)}]{Barker66p391}%
  \BibitemOpen
  \bibfield  {author} {\bibinfo {author} {\bibfnamefont {A.}~\bibnamefont
  {Barker~Jr}},\ }\href@noop {} {\bibfield  {journal} {\bibinfo  {journal}
  {Physical Review}\ }\textbf {\bibinfo {volume} {145}},\ \bibinfo {pages}
  {391} (\bibinfo {year} {1966})}\BibitemShut {NoStop}%
\bibitem [{\citenamefont {Wurfel}\ and\ \citenamefont
  {Batra}(1973)}]{Wurfel73p5126}%
  \BibitemOpen
  \bibfield  {author} {\bibinfo {author} {\bibfnamefont {P.}~\bibnamefont
  {Wurfel}}\ and\ \bibinfo {author} {\bibfnamefont {I.~P.}\ \bibnamefont
  {Batra}},\ }\href@noop {} {\bibfield  {journal} {\bibinfo  {journal} {Phys.
  Rev. B}\ }\textbf {\bibinfo {volume} {8}},\ \bibinfo {pages} {5126} (\bibinfo
  {year} {1973})}\BibitemShut {NoStop}%
\bibitem [{\citenamefont {Mehta}\ \emph {et~al.}(1973)\citenamefont {Mehta},
  \citenamefont {Silverman},\ and\ \citenamefont {Jacobs}}]{Mehta73p3379}%
  \BibitemOpen
  \bibfield  {author} {\bibinfo {author} {\bibfnamefont {R.~R.}\ \bibnamefont
  {Mehta}}, \bibinfo {author} {\bibfnamefont {B.~D.}\ \bibnamefont
  {Silverman}}, \ and\ \bibinfo {author} {\bibfnamefont {J.~T.}\ \bibnamefont
  {Jacobs}},\ }\href@noop {} {\bibfield  {journal} {\bibinfo  {journal} {J.
  Appl. Phys.}\ }\textbf {\bibinfo {volume} {44}},\ \bibinfo {pages} {3379}
  (\bibinfo {year} {1973})}\BibitemShut {NoStop}%
\bibitem [{\citenamefont {Berglund}\ and\ \citenamefont
  {Baer}(1967)}]{Berglund67p358}%
  \BibitemOpen
  \bibfield  {author} {\bibinfo {author} {\bibfnamefont {C.}~\bibnamefont
  {Berglund}}\ and\ \bibinfo {author} {\bibfnamefont {W.}~\bibnamefont
  {Baer}},\ }\href@noop {} {\bibfield  {journal} {\bibinfo  {journal} {Physical
  Review}\ }\textbf {\bibinfo {volume} {157}},\ \bibinfo {pages} {358}
  (\bibinfo {year} {1967})}\BibitemShut {NoStop}%
\bibitem [{\citenamefont {Boyeaux}\ and\ \citenamefont
  {Michel-Calendini}(1979)}]{Boyeaux79p545}%
  \BibitemOpen
  \bibfield  {author} {\bibinfo {author} {\bibfnamefont {J.}~\bibnamefont
  {Boyeaux}}\ and\ \bibinfo {author} {\bibfnamefont {F.}~\bibnamefont
  {Michel-Calendini}},\ }\href@noop {} {\bibfield  {journal} {\bibinfo
  {journal} {Journal of Physics C: Solid State Physics}\ }\textbf {\bibinfo
  {volume} {12}},\ \bibinfo {pages} {545} (\bibinfo {year} {1979})}\BibitemShut
  {NoStop}%
\bibitem [{\citenamefont {Fridkin}\ \emph {et~al.}(1973)\citenamefont
  {Fridkin}, \citenamefont {Grekov}, \citenamefont {Rodin}, \citenamefont
  {Savchenko},\ and\ \citenamefont {Volk}}]{Fridkin73p71}%
  \BibitemOpen
  \bibfield  {author} {\bibinfo {author} {\bibfnamefont {V.}~\bibnamefont
  {Fridkin}}, \bibinfo {author} {\bibfnamefont {A.}~\bibnamefont {Grekov}},
  \bibinfo {author} {\bibfnamefont {A.}~\bibnamefont {Rodin}}, \bibinfo
  {author} {\bibfnamefont {E.}~\bibnamefont {Savchenko}}, \ and\ \bibinfo
  {author} {\bibfnamefont {T.}~\bibnamefont {Volk}},\ }\href@noop {} {\bibfield
   {journal} {\bibinfo  {journal} {Ferroelectrics}\ }\textbf {\bibinfo {volume}
  {6}},\ \bibinfo {pages} {71} (\bibinfo {year} {1973})}\BibitemShut {NoStop}%
\bibitem [{\citenamefont {Vinetskii}\ \emph {et~al.}(1970)\citenamefont
  {Vinetskii}, \citenamefont {Itskovskii},\ and\ \citenamefont
  {Kukushkin}}]{Vinetskii70p23}%
  \BibitemOpen
  \bibfield  {author} {\bibinfo {author} {\bibfnamefont {V.}~\bibnamefont
  {Vinetskii}}, \bibinfo {author} {\bibfnamefont {M.}~\bibnamefont
  {Itskovskii}}, \ and\ \bibinfo {author} {\bibfnamefont {L.}~\bibnamefont
  {Kukushkin}},\ }\href@noop {} {\bibfield  {journal} {\bibinfo  {journal}
  {physica status solidi (b)}\ }\textbf {\bibinfo {volume} {39}},\ \bibinfo
  {pages} {K23} (\bibinfo {year} {1970})}\BibitemShut {NoStop}%
\bibitem [{\citenamefont {Burgess}\ and\ \citenamefont
  {Weiss}(1965)}]{burgess1965fluctuation}%
  \BibitemOpen
  \bibfield  {author} {\bibinfo {author} {\bibfnamefont {R.~E.}\ \bibnamefont
  {Burgess}}\ and\ \bibinfo {author} {\bibfnamefont {G.}~\bibnamefont
  {Weiss}},\ }\href@noop {} {\bibfield  {journal} {\bibinfo  {journal} {Physics
  Today}\ }\textbf {\bibinfo {volume} {18}},\ \bibinfo {pages} {60} (\bibinfo
  {year} {1965})}\BibitemShut {NoStop}%
\bibitem [{\citenamefont {Shockley}(1950)}]{shockley1953electrons}%
  \BibitemOpen
  \bibfield  {author} {\bibinfo {author} {\bibfnamefont {W.}~\bibnamefont
  {Shockley}},\ }\href@noop {} {\emph {\bibinfo {title} {Electrons and holes in
  semiconductors}}}\ (\bibinfo  {publisher} {D. Van Nostrand Company, Inc.},\
  \bibinfo {address} {New York},\ \bibinfo {year} {1950})\BibitemShut {NoStop}%
\bibitem [{\citenamefont {Kittel}\ and\ \citenamefont
  {McEuen}(1986)}]{kittel1986introduction}%
  \BibitemOpen
  \bibfield  {author} {\bibinfo {author} {\bibfnamefont {C.}~\bibnamefont
  {Kittel}}\ and\ \bibinfo {author} {\bibfnamefont {P.}~\bibnamefont
  {McEuen}},\ }\href@noop {} {\emph {\bibinfo {title} {Introduction to solid
  state physics}}},\ Vol.~\bibinfo {volume} {8}\ (\bibinfo  {publisher} {Wiley
  New York},\ \bibinfo {year} {1986})\BibitemShut {NoStop}%
\bibitem [{\citenamefont {Cardona}(1965)}]{Cardona65p651}%
  \BibitemOpen
  \bibfield  {author} {\bibinfo {author} {\bibfnamefont {M.}~\bibnamefont
  {Cardona}},\ }\href@noop {} {\bibfield  {journal} {\bibinfo  {journal}
  {Physical Review}\ }\textbf {\bibinfo {volume} {140}},\ \bibinfo {pages}
  {A651} (\bibinfo {year} {1965})}\BibitemShut {NoStop}%
\bibitem [{\citenamefont {Dietz}\ \emph {et~al.}(1997)\citenamefont {Dietz},
  \citenamefont {Schumacher}, \citenamefont {Waser}, \citenamefont {Streiffer},
  \citenamefont {Basceri},\ and\ \citenamefont {Kingon}}]{Dietz97p2359}%
  \BibitemOpen
  \bibfield  {author} {\bibinfo {author} {\bibfnamefont {G.}~\bibnamefont
  {Dietz}}, \bibinfo {author} {\bibfnamefont {M.}~\bibnamefont {Schumacher}},
  \bibinfo {author} {\bibfnamefont {R.}~\bibnamefont {Waser}}, \bibinfo
  {author} {\bibfnamefont {S.}~\bibnamefont {Streiffer}}, \bibinfo {author}
  {\bibfnamefont {C.}~\bibnamefont {Basceri}}, \ and\ \bibinfo {author}
  {\bibfnamefont {A.}~\bibnamefont {Kingon}},\ }\href@noop {} {\bibfield
  {journal} {\bibinfo  {journal} {Journal of applied physics}\ }\textbf
  {\bibinfo {volume} {82}},\ \bibinfo {pages} {2359} (\bibinfo {year}
  {1997})}\BibitemShut {NoStop}%
\bibitem [{\citenamefont {Gruverman}\ \emph {et~al.}(2009)\citenamefont
  {Gruverman}, \citenamefont {Wu}, \citenamefont {Lu}, \citenamefont {Wang},
  \citenamefont {Jang}, \citenamefont {Folkman}, \citenamefont {Zhuravlev},
  \citenamefont {Felker}, \citenamefont {Rzchowski}, \citenamefont {Eom} \emph
  {et~al.}}]{Gruverman09p3539}%
  \BibitemOpen
  \bibfield  {author} {\bibinfo {author} {\bibfnamefont {A.}~\bibnamefont
  {Gruverman}}, \bibinfo {author} {\bibfnamefont {D.}~\bibnamefont {Wu}},
  \bibinfo {author} {\bibfnamefont {H.}~\bibnamefont {Lu}}, \bibinfo {author}
  {\bibfnamefont {Y.}~\bibnamefont {Wang}}, \bibinfo {author} {\bibfnamefont
  {H.}~\bibnamefont {Jang}}, \bibinfo {author} {\bibfnamefont {C.}~\bibnamefont
  {Folkman}}, \bibinfo {author} {\bibfnamefont {M.~Y.}\ \bibnamefont
  {Zhuravlev}}, \bibinfo {author} {\bibfnamefont {D.}~\bibnamefont {Felker}},
  \bibinfo {author} {\bibfnamefont {M.}~\bibnamefont {Rzchowski}}, \bibinfo
  {author} {\bibfnamefont {C.-B.}\ \bibnamefont {Eom}},  \emph {et~al.},\
  }\href@noop {} {\bibfield  {journal} {\bibinfo  {journal} {Nano Lett.}\
  }\textbf {\bibinfo {volume} {9}},\ \bibinfo {pages} {3539} (\bibinfo {year}
  {2009})}\BibitemShut {NoStop}%
\bibitem [{\citenamefont {Garcia}\ \emph {et~al.}(2009)\citenamefont {Garcia},
  \citenamefont {Fusil}, \citenamefont {Bouzehouane}, \citenamefont
  {Enouz-Vedrenne}, \citenamefont {Mathur}, \citenamefont {Barthelemy},\ and\
  \citenamefont {Bibes}}]{Garcia09p81}%
  \BibitemOpen
  \bibfield  {author} {\bibinfo {author} {\bibfnamefont {V.}~\bibnamefont
  {Garcia}}, \bibinfo {author} {\bibfnamefont {S.}~\bibnamefont {Fusil}},
  \bibinfo {author} {\bibfnamefont {K.}~\bibnamefont {Bouzehouane}}, \bibinfo
  {author} {\bibfnamefont {S.}~\bibnamefont {Enouz-Vedrenne}}, \bibinfo
  {author} {\bibfnamefont {N.~D.}\ \bibnamefont {Mathur}}, \bibinfo {author}
  {\bibfnamefont {A.}~\bibnamefont {Barthelemy}}, \ and\ \bibinfo {author}
  {\bibfnamefont {M.}~\bibnamefont {Bibes}},\ }\href@noop {} {\bibfield
  {journal} {\bibinfo  {journal} {Nature}\ }\textbf {\bibinfo {volume} {460}},\
  \bibinfo {pages} {81} (\bibinfo {year} {2009})}\BibitemShut {NoStop}%
\end{thebibliography}%

\appendix
\section{Derivation of equation (4)}
\begin{equation*}
\begin{aligned}
&\phi_{2p}\left(\textbf{\emph{k}}\right)=\frac{1}{\left(2\pi\right)^{3/2}}\int{e^{-i\textbf{\emph{k}}\cdot\textbf{\emph{r}}}}\phi_{2p}\left(\textbf{\emph{r}}\right)d^3{\textbf{\emph{r}}} \\
&=\sqrt{\frac{1}{8\pi^3}}\int\int_{0}^{\infty}4\pi\sum_{l=0}^{\infty}\sum_{m=-l}^{l}\left(-i\right)^lj_l\left(kr\right)  \\
& \qquad  \qquad Y_{lm}^{*}\left(\theta_r,\phi_r\right)Y_{lm}\left(\theta_k,\phi_k\right)R\left(r\right)Y_{10}\left(\theta_r,\phi_r\right){r}^2drd{\Omega} \\ 
&=-\sqrt{\frac{1}{8\pi^3}}Y_{10}\left(\theta_k,\phi_k\right)\int_{0}^{\infty}4{\pi}ij_1\left(kr\right)R\left(r\right){r}^2dr  \\
&=-\sqrt{\frac{2}{\pi}}Y_{10}\left(\theta_k,\phi_k\right)\int_{0}^{\infty}i\left[\frac{\sin{\left(kr\right)}}{k^2r^2}-\frac{\cos{\left(kr\right)}}{kr}\right]R\left(r\right){r}^2dr  \\
&=-\sqrt{\frac{2}{\pi}}Y_{10}\left(\theta_k,\phi_k\right)\int_{0}^{\infty}i\left[\frac{e^{ikr}-e^{-ikr}}{2k^2i}-\frac{r\left(e^{ikr}+e^{-ikr}\right)}{2k}\right]   \\
& \qquad  \qquad  \qquad  \qquad \qquad  \qquad  \qquad  \qquad \qquad  \qquad R\left(r\right)dr \\
&=-\sqrt{\frac{2}{\pi}}\frac{c_1^{\prime}}{2k^2}Y_{10}\left(\theta_k,\phi_k\right)\int_{0}^{\infty}{r}\left(e^{-z_1{r}+ikr}-e^{-z_1{r}-ikr}\right)dr \\
&-\sqrt{\frac{2}{\pi}}\frac{c_2^{\prime}}{2k^2}Y_{10}\left(\theta_k,\phi_k\right)\int_{0}^{\infty}{r}\left(e^{-z_2{r}+ikr}-e^{-z_2{r}-ikr}\right)dr \\
&+\sqrt{\frac{2}{\pi}}\frac{c_1^{\prime}i}{2k}Y_{10}\left(\theta_k,\phi_k\right)\int_{0}^{\infty}{r^2}\left(e^{-z_1{r}+ikr}+e^{-z_1{r}-ikr}\right)dr \\
&+\sqrt{\frac{2}{\pi}}\frac{c_2^{\prime}i}{2k}Y_{10}\left(\theta_k,\phi_k\right)\int_{0}^{\infty}{r^2}\left(e^{-z_2{r}+ikr}+e^{-z_2{r}-ikr}\right)dr \\
\end{aligned}
\end{equation*}

\begin{equation*}
\begin{aligned}
&=-\sqrt{\frac{2}{\pi}}\frac{c_1^{\prime}}{2k^2}Y_{10}\left(\theta_k,\phi_k\right)\left[\frac{1}{\left(z_1-ik\right)^2}-\frac{1}{\left(z_1+ik\right)^2}\right] \\
&-\sqrt{\frac{2}{\pi}}\frac{c_2^{\prime}}{2k^2}Y_{10}\left(\theta_k,\phi_k\right)\left[\frac{1}{\left(z_2-ik\right)^2}-\frac{1}{\left(z_2+ik\right)^2}\right] \\
&+\sqrt{\frac{2}{\pi}}\frac{c_1^{\prime}i}{2k}Y_{10}\left(\theta_k,\phi_k\right)\left[\frac{2}{\left(z_1-ik\right)^3}+\frac{2}{\left(z_1+ik\right)^3}\right] \\
&+\sqrt{\frac{2}{\pi}}\frac{c_2^{\prime}i}{2k}Y_{10}\left(\theta_k,\phi_k\right)\left[\frac{2}{\left(z_2-ik\right)^3}+\frac{2}{\left(z_2+ik\right)^3}\right] \\
&=\sqrt{\frac{8}{\pi}}Y_{10}\left(\theta_k,\phi_k\right)\left[\frac{c_1^{\prime}\left(z_1^3-3z_1k^2\right)i}{k\left(z_1^2+k^2\right)^3}+\frac{c_2^{\prime}\left(z_2^3-3z_2k^2\right)i}{k\left(z_2^2+k^2\right)^3}\right.                \\
&\left.-\frac{c_1^{\prime}z_1i}{k\left(z_1^2+k^2\right)^2}-\frac{c_2^{\prime}z_2i}{k\left(z_2^2+k^2\right)^2}\right]   \\
&=-\sqrt{\frac{8}{\pi}}4iY_{10}\left(\theta_k,\phi_k\right)\left[\frac{c_1^{\prime}z_1k}{\left(z_1^2+k^2\right)^3}+\frac{c_2^{\prime}z_2k}{\left(z_2^2+k^2\right)^3}\right] \\
\end{aligned}
\end{equation*}

\section{Derivation of equation (8)}
The derivation follows the idea in Ref.~\cite{Mcquarrie1997physical}.
We consider that case that an electron with the wave vector $k$ penetrates a BaTiO$_3$ (BTO) nanowire.
\begin{figure}[htbp]
\includegraphics[width=8.0cm]{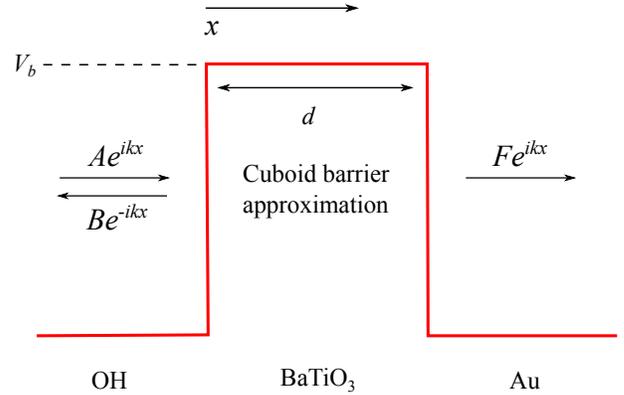}
\caption{Energy barrier diagram.}\label{a1}
\end{figure}

As shown in FIG.~\ref{a1}, The length of the penetration path is $d$. 
The energy of the electron is $E\left(\textbf{\emph{k}}\right)$ and the energy of the BaTiO$_3$ conduction band is $V_2$.
In the region of OH, the wave function is the plane wave including incident part and reflection part
\begin{equation} 
\varphi_1\left(x\right)=Ae^{ikx}+Be^{-ikx} \quad\left(x<0\right)
\end{equation}
\begin{equation} 
k=\left(\frac{2m_0E}{\hbar^2}\right)^{1/2}
\end{equation}
In the BaTiO$_3$ region, the Schr\"{o}dinger equation is 
\begin{equation} 
-\frac{\hbar^2}{2m_0}\frac{d^2\varphi\left(x\right)}{dx^2}+V\left(x\right)=E_{BTO}\varphi\left(x\right)
\end{equation} 
When we consider the electrons around the conduction band, we can write the potential energy term into effective mass
\begin{equation} 
-\frac{\hbar^2}{2m^*}\frac{d^2\varphi\left(x\right)}{dx^2}=E_{BTO}\varphi\left(x\right)
\end{equation} 
\begin{equation}
E_{BTO}=V_2-E\left(\textbf{\emph{k}}\right)
\end{equation}

\noindent where $m^*$ is the effective mass. The solution of the above Schr\"{o}dinger equation is 
\begin{equation}
\varphi_2\left(x\right)=Ce^{k^{'}x}+De^{-k^{'}x}\quad\left(0<x<d\right)
\end{equation}
\begin{equation}
k^{'}=\left(\frac{2m^*\left(V_2-E\right)}{\hbar^2}\right)^{1/2}
\end{equation}

With the analysis above, wave function in each region is summarized as
\begin{equation}
\left\{ \begin{array}{rl}
& \varphi_1\left(x\right)=Ae^{ikx}+Be^{-ikx}            \qquad      \mbox{$x<0$ }         \\
& \\
& \varphi_2\left(x\right)=Ce^{k^{'}x}+De^{-k^{'}x}      \qquad      \mbox{$0<x<d$ }      \\
& \\
& \varphi_3\left(x\right)=Fe^{ikx}            \qquad    \qquad       \qquad    \   \mbox{$x>d$ }
       \end{array} \right.
\end{equation}

The tunneling probability is given by
\begin{equation}
P=\frac{\left|F\right|^2}{\left|A\right|^2}
\end{equation}

$\varphi\left(x\right)$ and ${d\varphi\left(x\right)}/{dx}$ must be continuous at the boundaries.

\begin{equation}
A+B=C+D
\end{equation}

\begin{equation}
ik\left(A-B\right)=k^{'}\left(C-D\right)
\end{equation}

\begin{equation}
Ce^{k^{'}d}+De^{-k^{'}d}=Fe^{ikd}
\end{equation}

\begin{equation}
k^{'}Ce^{k^{'}d}-k^{'}De^{-k^{'}d}=ikFe^{ikd}
\end{equation}

From equation (B12) and (B13) we have,

\begin{equation}
C=\frac{\left(k^{'}+ik\right)Fe^{ikd}}{2k^{'}e^{k^{'}d}}
\end{equation}

\begin{equation}
D=\frac{\left(k^{'}-ik\right)Fe^{ikd}}{2k^{'}e^{-k^{'}d}}
\end{equation}

From (B10)$\times{ik}$+(B11), we have

\begin{equation}
A=\frac{\left(ik+k^{'}\right)C+\left(ik-k^{'}\right)D}{2ik}
\end{equation}
 
Substitute (B14) and (B15) into (B16) gives

\begin{equation}
A=\frac{Fe^{ikd}}{4ikk^{'}}\left[\left(ik+k^{'}\right)^2e^{-k^{'}d}-\left(ik-k^{'}\right)^2e^{k^{'}d}\right] 
\end{equation}

\begin{equation}
\frac{F}{A}=\frac{4ikk^{'}e^{-ikd}}{2\left(k^2-{k^{'}}^2\right)\sinh{k^{'}d}+4ikk^{'}\cosh{k^{'}d}}
\end{equation}

\begin{equation}
P=\frac{\left|F\right|^2}{\left|A\right|^2}=\frac{16k^2{k^{'}}^2}{4\left({k^{'}}^2-k^2\right)^2\sinh^2{k^{'}d}+16k^2{k^{'}}^2\cosh^2{k^{'}d}}
\end{equation}

Since $\cosh^2{k^{'}d}=1+\sinh^2{k^{'}d}$,

\begin{equation}
\begin{aligned}
P&=\frac{16k^2{k^{'}}^2}{4\left({k^{'}}^2k^2\right)^2\sinh^2{k^{'}d}+16k^2{k^{'}}^2}  \\
&=\cfrac{4}{4+
\cfrac{\left(k^2+{k^{'}}^2\right)^2}{k^2{k^{'}}^2}\sinh^2{k^{'}d}}
\end{aligned}
\end{equation}

Therefore,
\begin{multline}
P=
\cfrac{4}{4+
\cfrac{\left[m_0E+m^*\left(V_2-E\right)\right]^2}{m_0m^*E\left(V_2-E\right)}\sinh^2\left[\cfrac{2m^*d^2\left(V_2-E\right)}{\hbar^2}\right]^{1/2}}
\end{multline}

\end{document}